\documentclass[11pt]{article}
\usepackage[a4paper,margin=1in]{geometry}
\usepackage[utf8]{inputenc}
\usepackage[T1]{fontenc}
\usepackage{lmodern}
\usepackage{microtype}
\usepackage{graphicx}
\usepackage{multirow}
\usepackage{amsmath,amssymb,amsfonts}
\usepackage{amsthm}
\usepackage{mathrsfs}
\usepackage{dsfont}
\usepackage[title]{appendix}
\usepackage{xcolor}
\usepackage{textcomp}
\usepackage{booktabs}
\usepackage{algorithm}
\usepackage{algorithmicx}
\usepackage{algpseudocode}
\usepackage{listings}
\usepackage[numbers,sort&compress]{natbib}
\usepackage{authblk}
\usepackage[hidelinks]{hyperref}
\theoremstyle{plain}

\theoremstyle{definition}

\theoremstyle{remark}

\providecommand{\keywords}[1]{\par\noindent\textbf{Keywords:} #1}

\begin{document}

\title{Aerial Wildfire Suppression Planning with a Hybrid CNN and Cellular Automaton Fire Model}
\author[1]{Ion Matei\thanks{Corresponding author: \texttt{imatei@fujitsu.com}}}
\author[1]{Maksym Zhenirovskyy}
\author[1]{Takuya Kurihana}
\author[1]{Rohit Vuppala}
\author[1]{Hon Yung Wong}
\affil[1]{Space Data Frontiers Research Center, Fujitsu Research of America, 4655 Great America Pkwy, Santa Clara, California, 95054, USA}
\date{}

\maketitle

\begin{abstract}
Aerial wildfire suppression requires decisions about when, where, and how to deploy limited aircraft. We present an intervention-design framework built on a frozen hybrid convolutional neural network and cellular automaton (CNN-CA) simulator trained jointly on six historical wildfires. First, we jointly optimize binary drop execution and continuous location and orientation, with aircraft-specific footprints, wind drift, and availability, turnaround, and grounded-day constraints. Second, we model water as an immediate transfer of burning probability to the unburned state and retardant as a persistent reduction in the fuel contribution to spread. Third, we remove drops by a rollout-verified backward elimination while limiting degradation in selected fire-performance metrics. Fourth, we evaluate fixed schedules under daily state sampling, spatially correlated ignition-transition error, and their joint construction. Fifth, we compare the optimizer with random, tactical, greedy, and derivative-free planners under shared fleet, drop-budget, and simulator-evaluation allowances, and measure computational scaling. A 2020 Bear Fire case study considers two objectives: total fire-affected area and protected-region exposure. The two-stage total-area schedule reduces deterministic terminal extent by $89.5\%$ relative to the simulator baseline with $1{,}111$ drops. In Setting A, the nominal schedule reduces mean terminal extent by $41.5\%$ under fitted ignition-transition error without daily state sampling; designing at retardant/water effectiveness $0.85/0.75$ raises this reduction to $60.5\%$ with $1{,}850$ drops. With joint state sampling and transition error, the corresponding reductions are $86.0\%$ and $90.8\%$. The nominal schedule is sensitive to small pose and effectiveness perturbations. The tactical heuristic is more economical at small evaluation allowances; the gradient planner achieves better objectives with more computation. All reductions are conditional on the frozen simulator and modeled suppression response. They are not estimates of historical acreage saved or evidence of operational validity.
\end{abstract}


\section{Introduction}

Wildfire suppression is a sequential decision problem in which terrain, fuel, weather, and intervention effects shape fire evolution. A planning method must connect decisions about when, where, and how aircraft intervene to future spread while enforcing resource and operational constraints.

Wildfire models follow several approaches. Semi-empirical models such as the Rothermel equations describe fire behavior through interpretable physical relations~\citep{rothermel1972mathematical}. Operational systems such as FARSITE and FlamMap propagate perimeters and support fire-behavior, treatment, and conditional burn-probability analyses~\citep{finney1998farsite,finney2002mtt,finney2006flammap}. Cellular automata provide a computationally efficient alternative based on local state-transition rules~\citep{alexandridis2008cellular,freire2019cellular}. Machine-learning and hybrid models learn spread from observations while retaining varying degrees of physical structure~\citep{zheng2017forest,jain2020review,shen2023differentiable}.

Differentiable wildfire simulators support gradient-based calibration, accelerated simulation, and reinforcement-learning environments~\citep{xia2025pytorchfire,cakir2025jaxwildfire,maksym2026}. Suppression planning has largely developed separately, addressing tactical containment~\citep{fried1996containment,finney2009containment}, resource allocation and initial attack~\citep{belval2015mixed,ntaimo2013initialattack,avci2024wildfire,granda2023decision}, aerial logistics~\citep{rodriguezveiga2018aerial}, and strategic scenarios~\citep{houtman2013allowing,riley2018framework}. Predictive tools also identify control opportunities and attack success~\citep{oconnor2017empirical,cardil2025initialattack}. These methods generally do not differentiate final wildfire outcomes with respect to individual drops' locations and orientations through a learned spatial fire model. This gap motivates our formulation.

We build on the hybrid CNN-CA model of \citet{maksym2026}. Its shared network weights and fuel embedding remain frozen during intervention optimization. The architecture, training, learned spatial parameters, and predictive evaluation belong to that study; this paper contributes the intervention layer.

This work makes five contributions.
\begin{enumerate}
    \item \textbf{First,} we optimize binary drop execution and continuous location and orientation through the frozen CNN-CA simulator.
    \item \textbf{Second,} we construct a differentiable rollout with aircraft-specific footprints, wind drift, feasibility constraints, and separate water and retardant effects.
    \item \textbf{Third,} we develop a two-stage procedure that optimizes fire performance and then removes drops by a backward elimination in which every removal is verified by a rollout. A separate sensitivity study compares nominal-effectiveness designs with designs optimized for weaker suppressant.
    \item \textbf{Fourth,} we evaluate fixed schedules under daily fire-state sampling, spatially correlated ignition-transition error, and their joint construction.
    \item \textbf{Fifth,} we benchmark against random, tactical, greedy, and derivative-free planners under a common simulator and objective, and measure scaling with horizon, fleet size, and raster resolution.
\end{enumerate}
We demonstrate the pipeline on Bear 2020 and report effects relative to the simulator baseline.

Section~\ref{sec:provenance} establishes model provenance and evaluation scope. Sections~\ref{sec:opt} and~\ref{sec:uq} present intervention optimization and uncertainty analysis. Section~\ref{sec:results} reports the experiments, and Section~\ref{sec:comparison} discusses their implications and limitations. The appendices give implementation details.

\section{Wildfire-Model Provenance and Evaluation Scope}
\label{sec:provenance}

The frozen model of \citet{maksym2026} combines a multiscale CNN, a learned fuel embedding, spatially varying cellular-automaton parameters, and three-state transitions. One shared set of network weights and fuel-embedding parameters was trained jointly on six wildfires: Bear 2020, Chimney 2016, Pier 2017, Brattain 2020, Buck 2017, and Ferguson 2018.

The first 10 days of each event form the data-assimilation window; the fitted parameters are then frozen for forward prediction. The study reports intersection over union (IoU) above $0.6$ on all post-assimilation forecast days for five of the six events. Bear's event-mean IoU is $0.74$ across the 20-day analysis horizon, with increasing false-positive spread along part of the eastern perimeter on later days~\citep{maksym2026}. The simulator thus provides multi-event predictive evidence but does not reproduce Bear's historical perimeter exactly.

Bear contributed to model training, so our intervention experiment is not a held-out test of wildfire-model generalization. It evaluates intervention design for one event using the frozen model. The model study's multi-event temporal forecasts and this paper's one-event optimization results address distinct questions.

The training perimeters reflect fire growth during real suppression. The learned dynamics therefore incorporate those suppression regimes~\citep{maksym2026}. Our \textit{simulator baseline} is a Bear Fire rollout without the additional synthetic drops, not a historically unsuppressed fire. No training datasets were available for such unsuppressed events.

\section{Gradient-Based Optimization of Aerial Fire Suppression Interventions}
\label{sec:opt}

We formulate aerial intervention planning as finite-horizon control through the frozen CNN-CA model. After defining the state, we present five steps: feasible decisions, differentiable footprints, material effects, fire-performance optimization, and drop elimination.

\subsection{State representation}

At time $t$, the fire state at raster cell \(x \in \Omega\) is
\(
\mathbf{p}_t(x)=\bigl(p_t^{U}(x),\,p_t^{B}(x),\,p_t^{R}(x)\bigr)
\).
The components are the unburned, burning, and burned probabilities, respectively, with
\(
p_t^{U}(x)+p_t^{B}(x)+p_t^{R}(x)=1.
\)
The fire-occupancy probability is
\(
P_t^{\mathrm{fire}}(x)=p_t^{B}(x)+p_t^{R}(x)=1-p_t^{U}(x)
\),
the probability that the cell is burning or already burned.

\subsection{Step 1: Enforce feasible drop decisions}
\label{sec:step1}

We evaluate each candidate plan with a forward wildfire rollout in which, at every step $t$, each aircraft $a$ either executes one drop or remains idle. Let $T$ be the number of simulation steps. Let $N_a$ be the number of aircraft. For every step $t \in \{0,\dots,T-1\}$ and aircraft $a \in \{1,\dots,N_a\}$, the optimizer maintains a scalar drop logit $D_{t,a}$ and a three-dimensional pose logit:
\[
\mathbf{z}_{t,a}=\bigl(z^{(y)}_{t,a}, z^{(x)}_{t,a}, z^{(\theta)}_{t,a}\bigr).
\]
The following maps produce valid raster coordinates and orientations:
\[
y_{t,a}=(n_y-1)\frac{\sin\!\bigl(z^{(y)}_{t,a}\bigr)+1}{2},
\qquad
x_{t,a}=(n_x-1)\frac{\sin\!\bigl(z^{(x)}_{t,a}\bigr)+1}{2},
\]
\[
\theta_{t,a}=\pi\,\frac{\sin\!\bigl(z^{(\theta)}_{t,a}\bigr)+1}{2}.
\]
Here, $(n_y,n_x)$ is the raster size.

We implement binary drop execution with a straight-through estimator (STE). Let
\[
p_{t,a}=\sigma(D_{t,a}),
\]
where $\sigma(\cdot)$ is the logistic sigmoid. A drop is feasible only when aircraft $a$ is available and not grounded. The indicator $c_{t,a}\in\{0,1\}$ represents cooldown-based availability. The indicator $g_{t,a}\in\{0,1\}$ represents the grounded-day gate. The gated soft decision is
\[
\tilde{d}_{t,a}=p_{t,a}c_{t,a}g_{t,a}.
\]
The forward binary decision is
\[
d^{\mathrm{bin}}_{t,a}=\mathbb{I}\!\left[\tilde{d}_{t,a}>\tau_{\mathrm{STE}}\right],
\]
with the threshold $\tau_{\mathrm{STE}}=0.5$ (Table~\ref{tab:opt_params}), so that an available, non-grounded aircraft drops exactly when its drop logit is positive. The differentiable decision used in the rollout is
\[
d_{t,a}
=
\tilde{d}_{t,a}
+
\operatorname{stopgrad}\!\left(d^{\mathrm{bin}}_{t,a}-\tilde{d}_{t,a}\right),
\]
where $\operatorname{stopgrad}$ is the identity in the forward pass and has derivative zero in the backward pass. The forward pass therefore uses binary drops. The backward pass differentiates through the soft decision. This surrogate is the link between discrete execution and gradient-based optimization.

An executed drop restarts the aircraft-specific turnaround timer. With step duration $h=24/S_{\mathrm{day}}$ hours and assumed turnaround $\tau_a^{\mathrm{turn}}$, the counter reset is $\Delta_a=\lceil\tau_a^{\mathrm{turn}}/h\rceil$. The implemented gate checks the incoming counter before decrementing it, so drops are separated by at least $\Delta_a+1$ steps. This adds one step beyond rounded-up turnaround and conservatively restricts capacity. The availability indicator $c_{t,a}$ remains zero until the counter expires; grounded days set $g_{t,a}=0$. All reported planners use this same rule, specified in \eqref{eq:gate}.

\subsection{Step 2: Construct differentiable drop footprints}
\label{sec:step2}

We correct each commanded drop center for free-fall drift in the local wind. For aircraft $a$ with drop height $H_a$,
\[
t^{\mathrm{fall}}_a=\sqrt{\frac{2H_a}{g}},
\]
where $g$ is gravitational acceleration. Let $V_t(\cdot)$ be wind speed, and let $\psi_t(\cdot)$ be wind direction. Let $\Delta x$ be the grid spacing. Wind sampled at the commanded release point gives the landing point:
\[
x^{\mathrm{land}}_{t,a}
=
x_{t,a}
+
\frac{V_t(x_{t,a},y_{t,a})\cos\psi_t(x_{t,a},y_{t,a})\,t^{\mathrm{fall}}_a}{\Delta x},
\]
\[
y^{\mathrm{land}}_{t,a}
=
y_{t,a}
-
\frac{V_t(x_{t,a},y_{t,a})\sin\psi_t(x_{t,a},y_{t,a})\,t^{\mathrm{fall}}_a}{\Delta x}.
\]

The footprint represents the ground pattern of one drop. Its form follows the drop-test literature. A constant-flow tank releases the payload $P_a$ at a flow rate $q_a$ while the aircraft moves at airspeed $v_a$, so the released mass is uniform along a segment of length
\[
\ell_a=\frac{v_a P_a}{q_a}
\]
with mass per unit length $q_a/v_a$; the pattern length is this distance plus a dispersion term~\citep{legendre2014drop}. Measured along-track distributions of constant-flow drops are flat-topped and nearly symmetric, with ramps at the two ends that arise from the rise and the cut-off of the flow~\citep{amorim2011validation}, and the cross-track distribution of the liquid on the ground is Gaussian~\citep{legendre2014drop}. We therefore model the coverage as a uniform segment convolved with a Gaussian dispersion of scale $s_a$ along the track and $\sigma^{\perp}_a$ across it. The three aircraft-specific scales $\ell_a$, $s_a$ and $\sigma^{\perp}_a$ are calibrated from the published drop-guide tables or taken from aircraft analogues, and divided by $\Delta x$ to express them in cells (Appendix~\ref{app:Optimization setup and parameters}).

Define
\[
\Delta x_{ij}=j-x^{\mathrm{land}}_{t,a},
\qquad
\Delta y_{ij}=i-y^{\mathrm{land}}_{t,a},
\]
and the along-track and cross-track coordinates of cell $(i,j)$,
\[
u_{ij}=\cos\theta_{t,a}\,\Delta x_{ij}-\sin\theta_{t,a}\,\Delta y_{ij},
\qquad
w_{ij}=\sin\theta_{t,a}\,\Delta x_{ij}+\cos\theta_{t,a}\,\Delta y_{ij},
\]
so that the track follows $(\cos\theta,-\sin\theta)$ in raster coordinates, with rows increasing southward. The dispersion scales are corrected for the variance of a unit raster cell,
\[
\tilde{s}_a^{\,2}=s_a^2+\tfrac{1}{12},
\qquad
(\tilde{\sigma}^{\perp}_a)^2=(\sigma^{\perp}_a)^2+\tfrac{1}{12},
\]
which gives a Gaussian approximation to cell-averaged coverage; exact convolution with a square cell is not Gaussian. This correction is significant because the calibrated lines are narrower than a cell ($\sigma^{\perp}_a$ between $0.14$ and $0.25$ cells): without it, point sampling produces nearly flat across-track placement gradients through much of a cell and sharp peaks near cell boundaries. The footprint is
\[
G_{t,a}(i,j)
=
\frac{1}{2}\left[
\operatorname{erf}\!\left(\frac{u_{ij}+\ell_a/2}{\sqrt{2}\,\tilde{s}_a}\right)
-
\operatorname{erf}\!\left(\frac{u_{ij}-\ell_a/2}{\sqrt{2}\,\tilde{s}_a}\right)
\right]
\exp\!\left(-\frac{w_{ij}^2}{2(\tilde{\sigma}^{\perp}_a)^2}\right),
\]
a plateau of length $\ell_a$ with error-function ends along the track and a Gaussian profile across it. It is differentiable in the landing point and in the orientation. It is normalized over a square window $W_{t,a}$ centered at the nearest in-domain landing cell, whose half-width is half the longest segment plus five times the largest cell-corrected dispersion, rounded up to an integer:
\[
\bar{G}_{t,a}(i,j)=\frac{G_{t,a}(i,j)}{\sum_{(i',j')\in W_{t,a}}G_{t,a}(i',j')}.
\]
The implementation clips the cross-track exponent to $[-50,0]$ and adds $10^{-12}$ to the denominator for numerical stability. Discrete normalization is necessary because the lines are narrower than the $30$\,m raster. For the S-2T ($\sigma^{\perp}=0.17$ cells), the point-sampled cross-track sum of the uncorrected Gaussian varies between $0.02$ and $1.0$ with sub-cell placement, and that of the cell-corrected Gaussian between $0.66$ and $1.02$; an analytic normalization constant would therefore make the delivered mass depend on placement. Normalizing the realized footprint conserves the deposited payload within the evaluation window, up to numerical regularization; out-of-domain material is then discarded.

The local window avoids evaluating negligible tails over the full raster. Its half-width is $37$ cells in this configuration, set by the $1.8$\,km segment of the largest aircraft. Its padding also retains the distinction between payload deposited inside and outside the domain: material outside the raster is discarded rather than redistributed to boundary cells.

Wind is sampled by nearest-neighbour lookup at the release point, making the drift correction piecewise constant in release coordinates. Placement gradients flow through the footprint's dependence on the commanded position; the sampled wind contributes no gradient within a cell. Bilinear wind sampling could smooth this correction.

The footprint is differentiable in location and orientation between lookup and window boundaries; clipping and saturation introduce further nonsmooth points or flat regions. Ordinary pose gradients propagate through the deposit, modified state, and subsequent fire transitions to the loss. Only binary execution uses the straight-through estimator.

\subsection{Step 3: Apply material-specific suppression}
\label{sec:step3}

Let $P_a$ be the payload of aircraft $a$. The deposited intervention effect is
\[
E_{t,a}(i,j)=\bar{G}_{t,a}(i,j)\,P_a\,\alpha_a\,d_{t,a}.
\]
The coefficient $\alpha_a$ is not a free parameter. It is fixed by the effectiveness $\eta$ of \eqref{eq:suppression_dose} together with the assumed reference coverage $c_{\mathrm{req}}$,
\[
\alpha_a=\frac{-\ln\!\left(1-\eta_{m(a)}\right)}{c_{\mathrm{req}}\,A_{\mathrm{cell}}},
\qquad
A_{\mathrm{cell}}=\frac{(\Delta x/0.3048)^{2}}{100},
\]
where $m(a)\in\{\mathrm{water},\mathrm{retardant}\}$ is the material carried and $A_{\mathrm{cell}}$ is the cell area in units of $100$\,ft$^{2}$ ($96.9$ at $\Delta x=30$\,m). Because $\bar{G}_{t,a}P_a/A_{\mathrm{cell}}$ is the coverage $c$ delivered to a cell, this gives $E_{t,a}=-\ln(1-\eta_{m(a)})\,c/c_{\mathrm{req}}$, so the suppression $1-e^{-E_{t,a}}$ applied below is exactly \eqref{eq:suppression_dose}. We assume $\eta=0.9$ for both materials. Appendix~\ref{app:Intervention Optimization} gives the published estimates that inform this value and discusses the water case.

We do not rescale $\alpha_a$ with the time step. Each drop deposits its dose once: water changes the current state, while retardant changes a persistent field. Turnaround is discretized separately with the additional step specified above. For an aircraft initially available and ungrounded throughout a day, a $0.5$\,h turnaround permits $8$ drops at $S_{\mathrm{day}}=15$ and $17$ at $S_{\mathrm{day}}=50$. The corresponding minimum spacings are $3.2$\,h and $1.44$\,h, so the discrete capacity depends materially on temporal resolution. The aggregate effects are
\[
E^{\mathrm{water}}_t(i,j)=\sum_{a \in \mathcal{A}_{\mathrm{water}}} E_{t,a}(i,j),
\qquad
E^{\mathrm{ret}}_t(i,j)=\sum_{a \in \mathcal{A}_{\mathrm{ret}}} E_{t,a}(i,j).
\]

Water changes the current burning probability:
\[
s^{\mathrm{water}}_t(i,j)=\exp\!\left(-\min\!\bigl(E^{\mathrm{water}}_t(i,j),50\bigr)\right),
\]
\[
\tilde{p}^B_t(i,j)=p^B_t(i,j)\,s^{\mathrm{water}}_t(i,j),
\]
\[
\tilde{p}^U_t(i,j)
=
\operatorname{clip}\!\left(
p^U_t(i,j)+p^B_t(i,j)\bigl(1-s^{\mathrm{water}}_t(i,j)\bigr),0,1
\right).
\]

Retardant creates a persistent fuel-suppression field $r_t(i,j)\in[0,1]$. We initialize it with $r_0(i,j)=1$:
\[
r_{t+1}(i,j)=r_t(i,j)\exp\!\left(-\min\!\bigl(E^{\mathrm{ret}}_t(i,j),50\bigr)\right),
\qquad
f^{\mathrm{eff}}_t(i,j)=f_t(i,j)\,r_{t+1}(i,j).
\]

The pretrained CNN supplies fixed cellular-automaton parameter maps for each day: a baseline spread rate, wind and slope sensitivities, an ignition-gain coefficient $\alpha_\gamma$, and a fuel factor $f$~\citep{maksym2026}; the ignition gain of a cell is $\gamma_t=\alpha_{\gamma}f^{\mathrm{eff}}_t$, so retardant reduces both the potential of a treated neighbour and the gain of a treated cell. From these, the heat exposure of a cell at a step is the burning probability of its eight neighbours weighted by directional spread potentials,
\[
\lambda_t(x)=\sum_{i=1}^{8} \tilde{p}^{B,(i)}_t(x)\,\phi^{(i)}_t(x),
\]
where $\tilde{p}^{B,(i)}_t(x)$ is the water-modified burning probability of neighbour $i$. The spread potential $\phi^{(i)}_t(x)$ combines the baseline rate, retardant-modified fuel factor $f^{\mathrm{eff}}_t$, wind, and slope. On burnable cells, ignition follows the Poisson form used in training; the implementation sets ignition to zero outside the fuel mask:
\[
p_{\mathrm{ignite},t}=1-\exp\!\bigl(-\gamma_t\lambda_t\bigr),
\]
and with $T_{\mathrm{burn}}$ the burn duration in steps, the three-state update is
\[
p_{\mathrm{cont}}=1-\frac{1}{T_{\mathrm{burn}}},
\qquad
p^{\mathrm{new}}_t=\tilde{p}^U_t\,p_{\mathrm{ignite},t},
\qquad
p^{\mathrm{burnout}}_t=\tilde{p}^B_t(1-p_{\mathrm{cont}}),
\]
\[
p^U_{t+1}=\operatorname{clip}\!\left(\tilde{p}^U_t-p^{\mathrm{new}}_t,0,1\right),
\qquad
p^B_{t+1}=\operatorname{clip}\!\left(\tilde{p}^B_t+p^{\mathrm{new}}_t-p^{\mathrm{burnout}}_t,0,1\right),
\]
\[
p^R_{t+1}=\operatorname{clip}\!\left(p^R_t+p^{\mathrm{burnout}}_t,0,1\right),
\qquad
P^{\mathrm{fire}}_t=1-p^U_t.
\]

The water update transfers burning probability to the unburned state and preserves the probability simplex. It is a modeling approximation: the inherited model has no extinguished, wet, or temporarily nonflammable state, and adding such a state with its recovery dynamics would improve the physical interpretation of water effects. Retardant instead acts on the fuel and persists through $r_t$ for the rest of the horizon.

To summarize, the rollout maps feasible binary decisions and continuous poses into material effects, then propagates those effects through the frozen fire model.

\subsection{Step 4: Optimize fire performance}

The primary objective contains four terms. Two terms measure fire exposure and terminal fire extent over the whole domain, one measures fire exposure inside designated protected regions, and the fourth represents intervention cost:
\begin{equation}
\mathcal{L}
=
\lambda_{\mathrm{burn}}\mathcal{L}_{\mathrm{burn}}
+
\lambda_{\mathrm{final}}\mathcal{L}_{\mathrm{final}}
+
\lambda_{\mathrm{prot}}\mathcal{L}_{\mathrm{prot}}
+
\lambda_{\mathrm{budget}}\mathcal{L}_{\mathrm{budget}}.
\label{eq:objective}
\end{equation}
The nonnegative user-defined weights select the objective. The terms measure time-integrated fire occupancy, terminal occupancy, protected-region exposure, and executed drops, respectively.

A zero weight removes a term. We study total fire-affected area and protected-region exposure; Table~\ref{tab:loss_weights} gives the weights, and Section~\ref{sec:scope_objective} explains the settings.

The sum $\sum_x P_t^{\mathrm{fire}}(x)$ is a probability-cell count; multiplying by cell area gives an expected physical area. We use
\begin{align}
\nonumber
\mathcal{L}_{\mathrm{burn}} &=
\frac{1}{T_{\mathrm{opt}}|\Omega|}
\sum_{t=t_0}^{T}\sum_{x\in\Omega} P_t^{\mathrm{fire}}(x), \\
\nonumber
\mathcal{L}_{\mathrm{final}} &=
\frac{1}{|\Omega|}\sum_{x\in\Omega} P_T^{\mathrm{fire}}(x), \\
\nonumber
\mathcal{L}_{\mathrm{budget}} &=
\sum_{t,a} d_{t,a}, \\
\nonumber
\mathcal{L}_{\mathrm{prot}} &=
\frac{1}{|\mathcal{R}|}\sum_{R\in\mathcal{R}}
\frac{1}{T_{\mathrm{opt}}|R|}
\sum_{t=t_0}^{T}\sum_{x\in R} P_t^{\mathrm{fire}}(x).
\end{align}
Here, \(T_{\mathrm{opt}}=T-t_0+1\) counts the scored time steps, from the first intervention at \(t_0\) to the final index \(T\). The grid contains \(|\Omega|\) cells. The variable \(d_{t,a}\in\{0,1\}\) records whether aircraft \(a\) executes a drop at time \(t\).

The set \(\mathcal{R}=\{R_1,\dots,R_{|\mathcal{R}|}\}\) contains axis-aligned rectangular protected regions, with \(|R|\) cells in region \(R\). Normalizing each region by its own area gives small and large regions equal weight. Thus, even extensive coverage need not make this term proportional to \(\mathcal{L}_{\mathrm{burn}}\). Region placement is a modeling input representing priorities such as settlements, evacuation corridors, or infrastructure. Table~\ref{tab:protected_regions} lists the Bear Fire regions.

The two stages follow an $\varepsilon$-constraint approach to fire damage and drop count: first seek low fire losses, then remove drops while bounding their degradation. The first-stage budget coefficient $\lambda_{\mathrm{budget}}=10^{-6}$ acts as a small regularizer. Adam's normalization can make its effect on logit updates appreciable, and it need not preserve the minimizer of the fire-only objective (Section~\ref{sec:scope_objective}). Step~5 returns a local minimum under single removals when it completes a pass with no accepted removal; reaching the rollout cap alone does not establish this property.

We optimize the drop and pose logits with Adam. Each epoch evaluates one rollout, differentiates the loss, and applies a clipped-gradient update; we retain the best parameters. Straight-through gradients are biased surrogates used as local search directions, not exact derivatives of binary execution.

\paragraph{Initialization.}
The first stage is a local method, so the result depends on the initialization. We initialize the three kinds of variables from the baseline rollout, the simulation without drops, as follows. Every drop logit starts at $D_{t,a}=-0.04$, so that $\sigma(D_{t,a})=0.49$ is slightly below the execution threshold: no drop is executed at first, and a drop is executed only after the fire gradient through the straight-through estimator raises its logit above the threshold. Each release position starts on the fire front of its step rather than at the centroid of the burning region: water aircraft are placed on dispersed local maxima of the burning probability $p^{B}_t$, retardant aircraft on dispersed maxima of the forecast ignition $\max(0,\,p^{B}_{t+S_{\mathrm{day}}/2}-p^{B}_t)$, with a $12$-cell exclusion radius so that simultaneous drops form a line, the placement rule of the tactical heuristic of Section~\ref{sec:planner_benchmark}. Each heading starts across the local wind at the release cell, $\theta_{t,a}=\psi+\pi/2$ modulo $\pi$, the rule that lays a retardant line across the direction of spread. Section~\ref{sec:scope_objective} compares this initialization with alternatives.

To improve stability over long horizons, the optimizer truncates one long-range gradient path through the burning-state carry:
\[
p^{B,\mathrm{carry}}_{t+1}
=
\operatorname{stopgrad}(p^B_{t+1}).
\]
This operation reduces gradient explosion. It preserves the main local path from a water drop to the loss. The truncation is a safeguard for the gradient computation, not part of the suppression physics. Retardant does not require the same truncation because its persistent fuel-suppression field evolves through a bounded exponential factor.

\subsection{Step 5: Prune unnecessary drops}
\label{sec:step5}

The second stage removes drops from the first-stage schedule while keeping selected fire metrics within reference tolerances. It is a backward elimination on the set of executed drops: a removal is accepted only if a rollout of the resulting schedule satisfies every constraint. Positions and orientations remain fixed at their first-stage values.

\paragraph{Constraints.}
Let $\ell_{k}^{\mathrm{ref}}$, $k\in\{\mathrm{burn},\mathrm{final},\mathrm{prot}\}$, be the raw losses from a new evaluation of the reference plan, and let $\rho\geq 1$ be a user-defined slack factor. The admissible thresholds are
\[
\tau_k = \max\!\left(\rho\,\ell_k^{\mathrm{ref}},\tau_{\min}\right),
\]
with an absolute floor $\tau_{\min}=10^{-4}$ that also covers near-zero protected exposure. A schedule is \emph{feasible} if $\ell_k\le\tau_k$ for every active constraint $k$: burn and terminal extent in Setting A, protected-region exposure in Setting B, with $\rho=1.02$ in both. Each constraint acts on a raw, unweighted loss and is held separately, so the tolerance is relative to the attained performance of the first stage rather than to its weighted objective.

\paragraph{Decision set.}
Let $u^{\mathrm{ref}}$ be the first-stage schedule, with drop logits $D^{\mathrm{ref}}_{t,a}$ and poses $\mathbf{z}^{\mathrm{ref}}_{t,a}$, and let
\[
\mathcal{E}=\{(t,a): d^{\mathrm{bin}}_{t,a}(u^{\mathrm{ref}})=1\}
\]
be the set of pairs at which it executes a drop, so that $|\mathcal{E}|=\mathcal{L}_{\mathrm{budget}}(u^{\mathrm{ref}})$; we write $N(u)=\mathcal{L}_{\mathrm{budget}}(u)$ for the executed-drop count of a schedule. For a subset $\mathcal{S}\subseteq\mathcal{E}$, let $u(\mathcal{S})$ denote the schedule with the poses of $u^{\mathrm{ref}}$ and drop logits
\[
D_{t,a}=\begin{cases}D^{\mathrm{ref}}_{t,a}, & (t,a)\in\mathcal{S},\\ -3, & \text{otherwise.}\end{cases}
\]
The decision set of the second stage is $\{u(\mathcal{S}):\mathcal{S}\subseteq\mathcal{E}\}$. This restriction is needed because of the turnaround timer. Let $\Delta_a$ be the cooldown reset value of aircraft $a$ in steps. Removing an executed drop $(t,a)$ leaves aircraft $a$ available at the steps $t<t'<t+\Delta_a+1$ that the timer had blocked, so a pair $(t',a)$ with $D^{\mathrm{ref}}_{t',a}>0$ that $u^{\mathrm{ref}}$ did not execute would execute in its place, and the drop count would not decrease. With $D_{t',a}=-3$ for every $(t',a)\notin\mathcal{E}$, no such pair executes, $N(u(\mathcal{S}))=|\mathcal{S}|$ for every $\mathcal{S}$, and the retained drops keep their timing.

\paragraph{Order of removals.}
The search starts at $\mathcal{S}=\mathcal{E}$ and proceeds in passes. Each pass first sorts the elements of $\mathcal{S}$. Let
\[
g_{t,a}=\frac{\partial}{\partial D_{t,a}}\sum_{k}\mathbf{1}_{k}\,\frac{\ell_k(u(\mathcal{S}))}{\tau_k}
\]
be the straight-through gradient of the sum of the active constrained losses, each divided by its threshold. Lowering $D_{t,a}$ changes that sum by about $-g_{t,a}$ per unit to first order, so the elements are sorted by $-g_{t,a}$ in increasing order, with $D^{\mathrm{ref}}_{t,a}$ increasing on ties. The gradient is computed once per pass and is used only to decide the order in which removals are tested. Whether a removal is accepted is decided by a rollout, never by the gradient.

\paragraph{Batches.}
The sorted elements are tested in batches $\mathcal{C}\subseteq\mathcal{S}$ of $b$ consecutive elements. One rollout evaluates $u(\mathcal{S}\setminus\mathcal{C})$. If it is feasible, $\mathcal{S}\leftarrow\mathcal{S}\setminus\mathcal{C}$ and $b\leftarrow\min(2b,256)$. Otherwise $\mathcal{C}$ is split into two halves that are tested in the same way, down to single elements, and $b\leftarrow\max(\lfloor b/2\rfloor,1)$ for the next batch. An element $(t,a)$ for which $u(\mathcal{S}\setminus\{(t,a)\})$ is infeasible stays in $\mathcal{S}$. The initial batch size is $b=32$. Large batches make the early passes inexpensive, when most removals succeed; halving makes the late passes exact, when most fail.

\paragraph{Passes and stopping.}
A pass tests the removal of every element of $\mathcal{S}$ once. Passes repeat until one pass removes nothing. At that point $u(\mathcal{S}\setminus\{(t,a)\})$ is infeasible for every $(t,a)\in\mathcal{S}$, so no single remaining drop can be removed within the tolerance. The stage also stops when an allowance of $40{,}000$ rollouts is exhausted; the case study did not reach this limit. The output is $u(\mathcal{S})$.

\paragraph{Guarantees and cost.}
Every accepted $\mathcal{S}$ is one whose rollout was feasible, so the output satisfies the constraints by construction. When the stage stops after a pass with no removal, the output is a local minimum of $N$ under single removals; termination at the rollout cap guarantees only feasibility. It is not a minimum drop count: two elements that are jointly removable but not individually, and exchanges that move a drop rather than remove it, are outside the search. The cost is the number of rollouts. A pass over $|\mathcal{S}|=n$ elements costs at most about $2n$ rollouts when most batches fail and substantially fewer when most succeed. Section~\ref{sec:results} reports the rollout counts of the case study.

\paragraph{Design margin.}
The elimination enforces tolerances on nominal fire metrics; it does not impose a bound on sensitivity to placement, effectiveness, or fire-state errors. A candidate design margin can be introduced by running both stages with the effectiveness in \eqref{eq:suppression_dose} set to design values $\eta^{\mathrm{d}}$ below the nominal $\eta$, then evaluating the schedule at the nominal value. This asks the optimizer to meet its objectives with weaker suppressant. It does not certify robustness, and the extra optimization effort and retained drops must be measured. Section~\ref{sec:design_eta} reports that comparison. Both stages of the main reported schedules use the nominal effectiveness.

Appendix~\ref{app:Optimization setup and parameters} lists the parameters used.

\section{State Sampling and Ignition-Transition Uncertainty}
\label{sec:uq}
The deterministic rollout gives one predicted evolution. Daily categorical state sampling represents stochastic variation within the specified transition model, while spatially correlated perturbations of ignition-transition probabilities represent model discrepancy. We evaluate these two sources separately and jointly. The ignition perturbations preserve the simplex, prohibit ignition when its transition probability is zero, and retain the absorbing burned state.

\subsection{Daily state sampling}

Intuitively, aleatoric uncertainty is the variation generated when a probabilistic fire state resolves into a categorical daily state. More precisely, we generate Monte Carlo (MC) trajectories by sampling a categorical state at each cell at day boundaries and propagating it through the deterministic within-day transitions used in training and optimization. Optimization itself propagates fractional probabilities across days. Daily sampling preserves within-day accumulation of ignition probability while introducing stochastic day-to-day variation. Let
\[
X_n^{(m)}(x)\in\{U,B,R\}
\]
be the realized state of cell $x$ on day $n$ in Monte Carlo sample $m\in\{1,\dots,N\}$. Its one-hot indicators are
\[
I_{n,m}^{U}(x)=\mathbb{I}\{X_n^{(m)}(x)=U\},\qquad
I_{n,m}^{B}(x)=\mathbb{I}\{X_n^{(m)}(x)=B\},\qquad
I_{n,m}^{R}(x)=\mathbb{I}\{X_n^{(m)}(x)=R\}.
\]
The corresponding marginal probability estimates are
\begin{equation*}
\hat{p}_n^{U}(x)=\frac{1}{N}\sum_{m=1}^{N} I_{n,m}^{U}(x),\qquad
\hat{p}_n^{B}(x)=\frac{1}{N}\sum_{m=1}^{N} I_{n,m}^{B}(x),\qquad
\hat{p}_n^{R}(x)=\frac{1}{N}\sum_{m=1}^{N} I_{n,m}^{R}(x),
\end{equation*}
so that
\[
\hat{P}_n^{\mathrm{fire}}(x)=\hat{p}_n^{B}(x)+\hat{p}_n^{R}(x)=1-\hat{p}_n^{U}(x).
\]

In the state-sampling-only ensemble, stochasticity enters only through the categorical draws at day boundaries. Let $S_{\mathrm{day}}$ be the number of micro-steps per day. Suppressing the member index $m$ for the within-day probabilities, let
\[
\mathbf{p}_{n,s}(x)=\bigl(p_{n,s}^{U}(x),\,p_{n,s}^{B}(x),\,p_{n,s}^{R}(x)\bigr),
\qquad s=0,\dots,S_{\mathrm{day}},
\]
be the within-day state. Each member starts from its one-hot daily state at $s=0$ and evolves deterministically:
\[
\mathbf{p}_{n,s+1}
=
F_{\Theta_n,\mathcal{E}_n}\!\left(\mathbf{p}_{n,s}\right),
\qquad s=0,\dots,S_{\mathrm{day}}-1.
\]
Here, $F_{\Theta_n,\mathcal{E}_n}$ is the cellular-automaton transition operator. The daily CNN maps $\Theta_n$ and environmental inputs $\mathcal{E}_n$ parameterize this operator. After the final micro-step, we sample the next state independently across cells conditional on their predicted probabilities:
\[
X_{n+1}^{(m)}(x)
\sim
\mathrm{Categorical}\!\left(
p_{n,S_{\mathrm{day}}}^{U}(x),\,
p_{n,S_{\mathrm{day}}}^{B}(x),\,
p_{n,S_{\mathrm{day}}}^{R}(x)
\right).
\]

The baseline applies no additional suppression. Its initial state is
\[
p_0^{U}(x)=\mathbb{I}\{X_0(x)=U\},\qquad
p_0^{B}(x)=\mathbb{I}\{X_0(x)=B\},\qquad
p_0^{R}(x)=0.
\]
The intervention case uses the same sampling scheme and a fixed precomputed schedule of executed drops and poses at every simulation step $t$:
\[
d_{t,a}\in\{0,1\},\qquad (y_{t,a},x_{t,a},\theta_{t,a}),\qquad t=0,\dots,T-1,
\]
Thus, the aleatoric ensemble is conditional on the chosen intervention schedule.

The sampled trajectories yield marginal state probabilities and trajectory-level metrics. Let $n_{\mathrm{days}}=T/S_{\mathrm{day}}$ be the number of simulated days ($15$ here). The final fire extent is
\[
A_{n_{\mathrm{days}}}^{(m)}=\sum_{x\in\Omega}\mathbb{I}\{X_{n_{\mathrm{days}}}^{(m)}(x)\neq U\}.
\]
The daily fire exposure is
\[
A_n^{(m)}=\sum_{x\in\Omega}\mathbb{I}\{X_n^{(m)}(x)\in\{B,R\}\}.
\]
We store ensemble mean and standard-deviation fields and each member's daily whole-domain and protected-region totals. Daily CNN maps are precomputed, and trajectories are vectorized and processed in batches.

\subsection{Spatially correlated ignition-transition error}
\label{sec:prediction_error_noise_supp}

We represent model discrepancy by perturbing ignition transitions while retaining the CA's allowed state transfers. Let $q_t(x)$ denote the ignition probability after accounting for the burnable mask, neighbouring burning probabilities, and the effective fuel factor, including retardant. The two outcomes for an unburned cell have probabilities $(q_t,1-q_t)$. For $0<q_t<1$, their isometric log-ratio (ILR) coordinate~\citep{egozcue2003isometric} is
\begin{equation}
z_t(x)=\frac{1}{\sqrt{2}}\log\frac{q_t(x)}{1-q_t(x)}.
\end{equation}
At every step $t$ of day $n$, we add the same daily field $\epsilon_n(x)$ to the current ignition coordinate $z_t(x)$. The resulting ignition probability is
\begin{equation}
\widetilde q_t(x)=\frac{q_t(x)\exp(\sqrt{2}\epsilon_n(x))}
{1-q_t(x)+q_t(x)\exp(\sqrt{2}\epsilon_n(x))},\qquad n=\lfloor t/S_{\mathrm{day}}\rfloor.
\label{eq:transition_ilr}
\end{equation}
We retain the endpoints $q_t=0$ and $q_t=1$ exactly, without replacing zeros by a positive constant. A stable odds calculation avoids logarithms at these endpoints and exponential overflow.

With the original burnout probability $\beta=1/T_{\mathrm{burn}}$, propagation becomes
\begin{align}
p^U_{t+1}&=p^U_t(1-\widetilde q_t),\nonumber\\
p^B_{t+1}&=p^B_t(1-\beta)+p^U_t\widetilde q_t,\label{eq:transition_uq_update}\\
p^R_{t+1}&=p^R_t+\beta p^B_t.\nonumber
\end{align}
For interventions, the water update precedes these equations and retardant enters $q_t$ through the effective fuel factor. Thus uncertainty changes the rate of an allowed transition; it never transfers burned probability back to an earlier state. Each update preserves nonnegativity and unit total probability, and $p^R$ cannot decrease. A cell with $q_t=0$ remains nonignitable at that step, while a fully burned cell remains burned. A cell may ignite later if neighbouring spread makes $q_t$ positive. Spotting or other external ignition sources would require an explicit transition mechanism. Initial ignition is restricted to the burnable mask; no initially burning cells are excluded in the reported Bear initialization.

We use
\begin{equation}
\epsilon_n(x)=\mu+\sigma Z_n(x),\qquad
\operatorname{Cov}(Z_n(x),Z_n(x'))\simeq\exp(-\|x-x'\|/\ell).
\end{equation}
A fresh zero-mean, unit-variance Gaussian field $Z_n$ is drawn at the start of each day and held fixed for its $S_{\mathrm{day}}=50$ microsteps. Successive daily fields are independent. We sample $Z_n$ by filtering Gaussian white noise in the Fourier domain on a padded periodic grid of dimensions at least twice the raster dimensions. Negative eigenvalues of the discretized exponential kernel are clipped to zero and the spectrum is normalized to unit marginal variance. This specifies approximate spatial covariance. Burn duration is held fixed because the observed fire masks do not separately identify burning and burned states. In the transition-error-only ensemble, fractional state probabilities propagate across day boundaries without categorical sampling.

\paragraph{Fitting through simulated trajectories.}
The mean, standard deviation, and correlation length are selected by running the modified simulator, rather than reusing a covariance of cumulative state residuals as daily transition noise. We compare block averages of simulated $p^B+p^R$ with the observed affected fraction on $16\times16$-cell blocks ($480\times480$\,m). Both fields are restricted to the burnable mask; boundary blocks are zero-padded. This aggregation is a scoring scale, not a claim of observational resolution. No unobserved burning/burned split is assigned to the observations.

For day $n$, let $a_n^{(j)}\in\mathbb R^K$ be the simulated block fractions of member $j$ and $y_n$ the observed block fractions. We minimize the mean energy score~\citep{gneiting2007proper} over Days~1--10, using distinct ensemble-member pairs to estimate its second expectation:
\begin{equation}
\widehat{\mathrm{ES}}_n=
\frac{1}{m\sqrt K}\sum_{j=1}^m\|a_n^{(j)}-y_n\|_2
-\frac{1}{2m(m-1)\sqrt K}\sum_{j\ne k}\|a_n^{(j)}-a_n^{(k)}\|_2.
\label{eq:transition_energy_score}
\end{equation}
Lower scores are better. The search uses $m=32$ members and common random fields across candidates, with $\mu\in\{-2,-1.5,-1,-0.5,0\}$, $\sigma\in\{0.5,1,1.5,2,2.5\}$, and $\ell\in\{1,4,16,64\}$ cells. Five additional candidates use $\sigma=0$, including the original deterministic simulator. Selection uses only fitting-day scores. Days~11--15 provide a subsequent temporal evaluation with fresh random fields; they are not used to select parameters. Table~\ref{tab:grf_params} reports the selected values and scores. The CNN remains frozen throughout.

\begin{table}[ht]
\centering
\caption{Ignition-transition error calibration on Bear 2020 without categorical state sampling. Energy scores are lower-is-better spatial scores on block fractions, not area errors. Post-selection scores use 32 fresh members.}
\label{tab:grf_params}
\begin{tabular}{ll}
\toprule
Quantity & Value \\
\midrule
Selected ILR perturbation mean $\mu$ & $-0.5$ \\
Selected ILR perturbation standard deviation $\sigma$ & $1.0$ \\
Selected correlation length $\ell$ & $1$ cell ($30$\,m) \\
Fitting window; temporal evaluation & Days 1--10; Days 11--15 \\
Candidates; members per candidate & $105$; $32$ \\
Selection energy score (Days 1--10) & $0.0572$ \\
Post-selection score, Days 1--10 & $0.0562$ (deterministic: $0.0771$) \\
Post-selection score, Days 11--15 & $0.2065$ (deterministic: $0.3373$) \\
Final schedule-evaluation members & $300$ \\
\bottomrule
\end{tabular}
\end{table}

The fitted discrepancy combines spread-model error, observation error, and effects of historical suppression represented in the observed event. Daily independence and spatial stationarity are assumptions; the search does not identify a general temporal error law or establish transfer to other fires. The temporal evaluation uses the same fire as calibration, and its first ten days also informed CNN training. The resulting ensemble therefore represents a specified, empirically fitted transition-error model, rather than an independently validated operational forecast distribution.

\subsection{Joint state sampling and ignition-transition error}
\label{sec:joint_uq}

The joint experiment draws one ignition-error field at the start of each day, reuses it in \eqref{eq:transition_ilr} at all $50$ microsteps, and samples a categorical state at each cell at the end of the day. The next day starts from this sampled one-hot state and a newly drawn error field. Retardant accumulated on previous days persists through the reset, and the drop schedule remains fixed. The categorical sampler assigns zero probability to impossible outcomes, so an already burned cell remains burned and a nonburnable cell cannot become burning through sampling.

The sampled state can be interpreted as a perfectly observed realization within the simulated model. It is drawn from that member's predicted probabilities, rather than supplied by an external observation. This experiment tests the joint propagation of state sampling and transition error; it does not perform observation assimilation or feedback replanning. We retain the error parameters fitted without categorical resets (Table~\ref{tab:grf_params}) to isolate the effect of adding daily sampling. They are not refitted for the joint construction. A companion state-only control uses the same sampler with $\mu=\sigma=0$.

\subsection{Conditional evaluation of a fixed schedule}

We evaluate state sampling alone, ignition-transition error alone, and their joint construction with $300$ members each. All six fixed schedules use the same saved daily CNN parameter fields and are evaluated at $\eta=0.9$. Within each construction, baseline and intervention members share random draws: ignition-error fields where applicable and categorical sampling variates where applicable. The state-only and joint constructions also share their categorical variates. This pairing supports differences between individual members as well as comparisons of ensemble means.

For a terminal metric $M$, the reported percentage reduction is $100(1-\overline M_{\mathrm{intervention}}/\overline M_{\mathrm{baseline}})$, using the baseline of that same construction. It is not the mean of memberwise percentage reductions. Transition-error-only members contribute sums of fractional probabilities; state-only and joint members contribute counts of realized burning or burned cells after the daily sample. Protected-region metrics sum over the union of the three designated regions and differ from the time- and region-normalized optimization loss. Plotted bands show ensemble spread, not confidence intervals. Numerical checks at daily boundaries monitor normalization, nonnegativity, irreversible burned-state history, and absence of burning on nonburnable cells, before and, where applicable, after the categorical reset.

\section{Experimental Results}
\label{sec:results}

We describe the Bear Fire scenario, report results for the two objectives, and compare planner performance and computational scaling.

\subsection{Scope and data}

The case study is a hypothetical aerial-suppression scenario for Bear 2020, within the simulator-relative scope of Section~\ref{sec:provenance}.

The frozen model receives Bear-specific LANDFIRE terrain and fuel data~\citep{landfire}, daily aggregates of the ERA5-Land reanalysis wind~\citep{joaquin2019era5land}, and observed daily perimeters from \citet{xia2025pytorchfire_data}.

\paragraph{Consistency of the perimeter record with the incident reports.}
We compared the area of the daily perimeters in this dataset with the Bear Fire acreage in the incident updates~\citep{inciweb2020northcomplex}. The two are close at the end of the window: $12{,}457$ against $12{,}154$ acres on $4$ September, a difference of approximately $2\%$. Earlier, the dataset exceeds the reported area by a factor of $28$ on $20$ August, $2.2$ on $23$ August, $1.5$ on $28$ August and $1.1$ on $1$ September. Matching areas across dates suggests a two-to-four-day lead, but does not establish agreement in perimeter geometry or a known reporting delay. The dataset has one connected fire region each day; this alone does not exclude merged detections. Its fire maps derive from MODIS thermal observations at approximately $1$\,km native resolution~\citep{xia2025pytorchfire,xia2025pytorchfire_data}. Differences in observation resolution, processing, and incident mapping are possible explanations for the discrepancy, whose cause has not been established. Resampling onto the $30$\,m simulator grid does not establish perimeter accuracy at that scale. Because wind, grounding, and intervention timing are tied to calendar dates, the discrepancy can affect planning opportunities as well as apparent growth rates. Its effect on the optimized schedules remains unquantified; the reported reductions are conditional on the dataset and simulator baseline.

\subsection{Fleet selection}

The fleet is an assumption. We did not find a clear record of the aircraft that operated on the Bear Fire: the incident updates~\citep{inciweb2020northcomplex} and the siege report~\citep{calfire2020} do not list a fleet assigned to it. The fleet assumed here consists of aircraft types that are typically used in aerial firefighting, and we ask what such a fleet could have achieved. It comprises $21$ aircraft: 6 S-2T Turbo Trackers, 2 AT-802F Fire Boss aircraft, 4 BAe-146/Avro RJ85 aircraft, 2 MD-87s, 2 C-130 MAFFS aircraft, 2 DC-10s, 1 Boeing 747 Supertanker, and 2 CL-415 Super Scoopers. The types represent the single-engine, large and very large airtanker classes and the scooper class; among them, the Boeing 747 and the DC-10 were contracted by CAL FIRE from 21 August 2020 during the fire siege of that year~\citep{calfire2020}. The counts are our choice. We exclude helicopters. Table~\ref{tab:aircraft_params} lists the specifications.

Smoke restricted aviation in the simulated window. The update of $22$ August reports that ``smoke inhibited the use of aircraft again today'', and the update of $30$ August that the lack of visibility prevented aerial reconnaissance~\citep{inciweb2020northcomplex}; the siege report states, for the 2020 siege as a whole, that smoke conditions hampered aircraft over multiple days~\citep{calfire2020}. We represent these restrictions by grounded days. All aircraft are grounded on $21$ and $22$ August, day indices $2$ and $3$ counted from zero at $19$ August, in agreement with that update, whose wording covers both days. The additional grounding of the large and very large airtankers on $23$ August, a day of Red Flag conditions and pyrocumulus development in the same record, is an assumption. The optimizer enforces the grounded days directly.

\subsection{Intervention design}

\paragraph{What is optimized.}
\label{sec:scope_objective}
Table~\ref{tab:loss_weights} defines two settings of \eqref{eq:objective}.

\emph{Setting A, total fire-affected area.} The first stage uses $\lambda_{\mathrm{burn}}=\lambda_{\mathrm{final}}=50$, $\lambda_{\mathrm{prot}}=0$, and $\lambda_{\mathrm{budget}}=10^{-6}$. The budget term encourages deactivation of drops with little fire-performance benefit. Although small in the scalar objective, its gradient can affect logit updates through Adam's per-coordinate normalization; a small coefficient does not guarantee a negligible change in the optimization trajectory. Preliminary budget-weight controls used an earlier Gaussian-footprint model and motivated this choice. They do not establish its effect on the constant-flow footprint model reported here. The second stage uses the first-stage logits only to break ties when it sorts the removals.

\emph{Initialization.} Preliminary single-seed experiments with an earlier Gaussian-footprint model motivated the front placement and wind-informed headings of Section~\ref{sec:opt}. Those experiments compared positive and near-threshold drop logits, centroid and front placement, and random and wind-informed headings. Their losses and convergence rates are not ablations of the present constant-flow footprint model. The reported runs use front initialization, the budget weight above, a cap of $10{,}000$ epochs, and a patience of $2{,}000$ epochs without improvement. The second stage removes drops while keeping burn and terminal extent within $2\%$ of their first-stage values. A matched initialization and budget-weight study for the revised deposition model remains to be performed.

\emph{Setting B, protected-region exposure.} The first stage uses $\lambda_{\mathrm{prot}}=100$ and $\lambda_{\mathrm{final}}=1$, with the other weights zero. Protection is the priority; terminal extent remains a secondary objective. The three protected regions cover $65.2\%$ of the domain (Table~\ref{tab:protected_regions}). Setting A uses the initialization of Section~\ref{sec:opt}. Setting B starts from the Setting A first-stage schedule, with the budget weight at zero, to reuse its dispersed suppression schedule. Drops outside the protected regions can influence protection through subsequent spread, although their gradients may be weak or zero. The terminal term provides a signal outside the protected regions. If the warm start already attains a protection loss at or below the threshold floor $10^{-4}$, the first stage is skipped and the second stage starts from the Setting A schedule. The second stage constrains only protected exposure, with threshold $\max(1.02\,\ell_{\mathrm{prot}}^{\mathrm{ref}},10^{-4})$; it does not constrain total burn or extent.

The weights were selected iteratively to obtain the reported behavior; they are not calibrated quantities.

We evaluate each of the six design schedules through a deterministic rollout and the three ensemble constructions of Section~\ref{sec:uq}: state sampling alone, ignition-transition error alone, and their combination.

The deterministic and state-sampling ensemble means need not agree. Let $\mathbf I_n$ denote the random one-hot state field at day $n$. Categorical sampling preserves state probabilities in expectation at each day boundary, but subsequent nonlinear transitions generally satisfy
\[
\mathbb{E}\!\left[F_n(\mathbf I_n)\right]
\neq
F_n\!\left(\mathbb{E}[\mathbf I_n]\right),
\]
where $F_n$ propagates the state through all within-day steps of day $n$. The deterministic rollout propagates fractional probabilities; sampled trajectories restart each day from categorical states. This distinction can shift the ensemble mean without losing probability mass. Here state sampling gives a smaller mean terminal extent than deterministic propagation.

Tables~\ref{tab:opt_params}, \ref{tab:loss_weights} and~\ref{tab:protected_regions} in Appendix~\ref{app:Optimization setup and parameters} list the optimization parameters, the loss weights, and the protected regions.

\subsubsection{Design effectiveness}
\label{sec:design_eta}
This subsection compares design effectiveness values in both stages of Setting A, evaluating every resulting schedule at the nominal $\eta=0.9$ (Table~\ref{tab:design_eta}). The nominal design has the lowest deterministic fire objective, the fewest drops, and the largest reduction under daily state sampling. Under ignition-transition error alone, however, the $0.85/0.75$ design increases the mean terminal-extent reduction from $41.5\%$ to $60.5\%$, with $1{,}850$ rather than $1{,}111$ drops. Lower design effectiveness also reduces sensitivity in the tested small pose perturbations: adding random perturbations at scale $10^{-4}$ to all pose logits multiplies the nominal schedule's fire objective by $1.7$ on average over four draws and by $2.3$ at most, whereas the objective of the $0.85/0.75$ design changes by $0.4\%$. At perturbation scale $10^{-3}$ the mean factors are $4.4$ and $1.5$. These are logit perturbations; physical displacements depend on the sine decoding and the individual poses. The $0.85/0.75$ design uses $67\%$ more drops, takes $9{,}660$ first-stage epochs against $5{,}910$, and removes $30\%$ of its drops in the second stage against $59\%$. The $0.80/0.70$ design loses ten percentage points of deterministic extent reduction and has a lower state-sampling reduction, but its $57.1\%$ transition-error-only reduction exceeds the nominal design's. In the joint ensemble, the $0.85/0.75$ design also improves terminal-extent reduction, from $86.0\%$ to $90.8\%$, whereas the $0.80/0.70$ design gives $76.7\%$. These comparisons demonstrate reduced sensitivity to the tested perturbations for an alternative, more expensive schedule, rather than a general robustness guarantee.

\begin{table}[htp!]
\centering
\footnotesize
\caption{Setting A schedules obtained at three design effectiveness values, all evaluated at the nominal $\eta=0.9$. Mean-field reduction is the deterministic terminal-extent reduction. Pose is the factor by which a perturbation of $10^{-4}$ in all pose logits multiplies $J_{\mathrm{fire}}$, mean over four draws. The three ensemble columns give Day-15 whole-domain reductions: state sampling alone, transition error alone, and their joint construction. Each reduction compares $300$-member means with the baseline of the same construction. Cost is the first-stage epoch count and the second-stage rollout count.}
\label{tab:design_eta}
\setlength{\tabcolsep}{2.5pt}
\begin{tabular}{lrrrrrrrr}
\toprule
Design $\eta^{\mathrm{d}}_{\mathrm{ret}}/\eta^{\mathrm{d}}_{\mathrm{w}}$ & Drops & $J_{\mathrm{fire}}$ & Mean-field & Pose & \shortstack{State\\only} & \shortstack{Transition\\only} & Joint & Cost \\
\midrule
0.90 / 0.90 & 1,111 & 2.77 & 89.5\% & 1.7 & 96.4\% & 41.5\% & 86.0\% & 5,910; 4,004 \\
0.85 / 0.75 & 1,850 & 4.11 & 84.3\% & 1.00 & 95.7\% & 60.5\% & 90.8\% & 9,660; 6,018 \\
0.80 / 0.70 & 1,757 & 5.56 & 79.5\% & 1.00 & 79.6\% & 57.1\% & 76.7\% & 3,660; 7,845 \\
\bottomrule
\end{tabular}
\end{table}

The nominal first-stage schedule is also sensitive to the effectiveness at which it is evaluated (Table~\ref{tab:eta_sens}): changing $(\eta_{\mathrm{ret}},\eta_{\mathrm{w}})$ from $(0.90,0.90)$ to $(0.89,0.88)$ multiplies terminal extent by $1.4$, to $(0.88,0.88)$ by $2.7$, and to $(0.85,0.80)$ by $3.5$. This schedule has not been pruned, so elimination cannot explain its effectiveness sensitivity. The pose and effectiveness tests concern different schedules and perturbations; they do not separate containment thresholds, raster/deposition effects, and pruning as causes. The large changes limit the robustness of the nominal result. The $2\%$ pruning tolerance bounds changes in selected deterministic losses during deletion, not changes caused by perturbed inputs. Robust open-loop design, robustness-aware pruning, and receding-horizon re-optimization are possible responses that have not been compared here.

\begin{table}[htp!]
\centering
\footnotesize
\caption{Sensitivity of the Setting A first-stage schedule ($2{,}727$ drops, obtained at $\eta=0.9$) to the effectiveness at which it is evaluated. Terminal extent in probability-cells; the baseline is $167{,}865$.}
\label{tab:eta_sens}
\begin{tabular}{lrr}
\toprule
$\eta_{\mathrm{ret}}$ ($\eta_{\mathrm{w}}$) & Terminal extent & Relative to $0.9$ \\
\midrule
0.90 (0.90) & 17,340 & 1.0 \\
0.895 (0.895) & 20,270 & 1.2 \\
0.89 (0.88) & 24,125 & 1.4 \\
0.88 (0.88) & 47,300 & 2.7 \\
0.86 (0.83) & 56,982 & 3.3 \\
0.85 (0.80) & 59,910 & 3.5 \\
0.70 (0.60) & 115,173 & 6.6 \\
\bottomrule
\end{tabular}
\end{table}

\subsubsection{Setting A: total fire-affected area}
\label{sec:results_a}
\begin{figure}[htp!]
    \centering
    \includegraphics[width=0.97\textwidth]{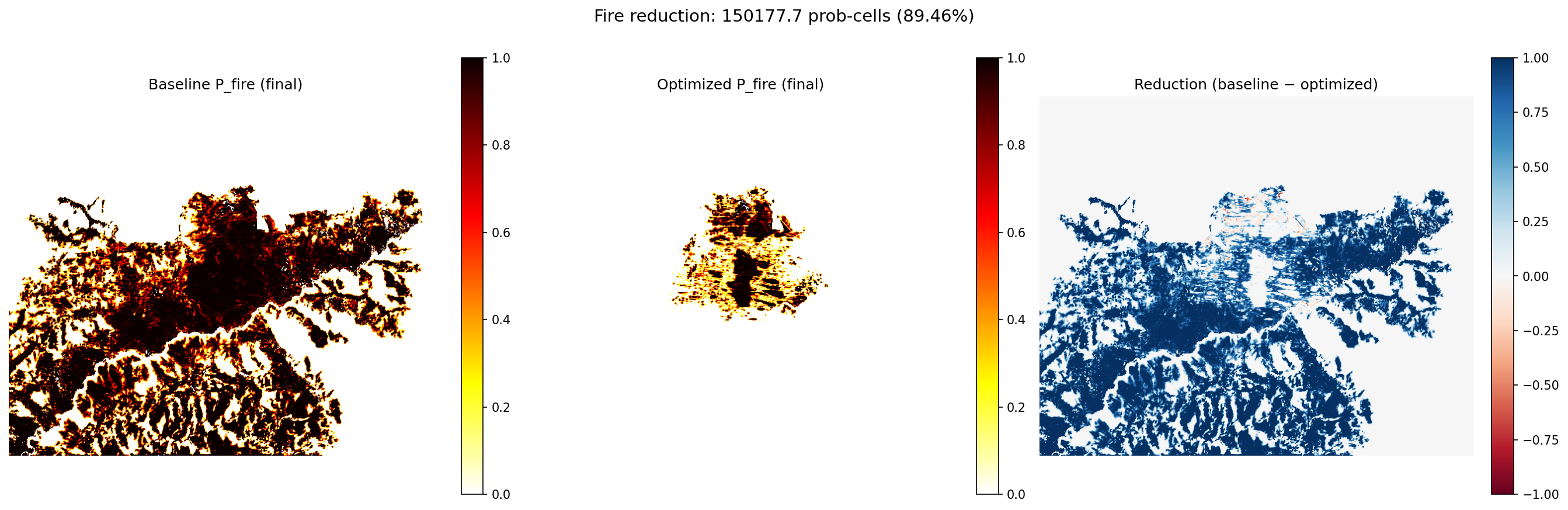}
    \caption{Setting A. Final fire-occupancy maps for the baseline and the optimized schedule after the backward elimination, and their difference. The deterministic rollout reduces the terminal fire-affected area by $89.5\%$.}
    \label{fig:fire_difference_map_area_min}
\end{figure}
\begin{figure}[htp!]
    \centering
    \includegraphics[width=0.97\textwidth]{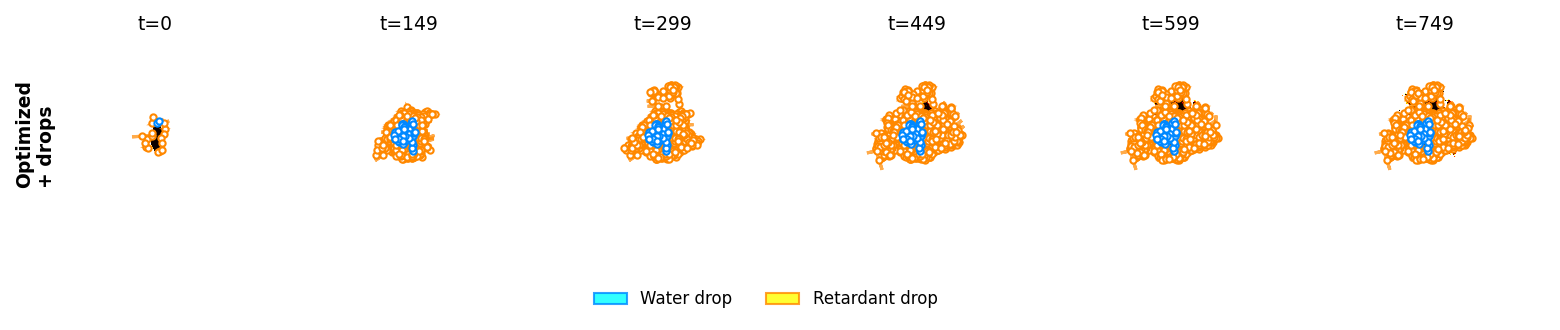}
    \caption{Setting A. Spatial distribution of the optimized aerial drops over time, after the backward elimination: $1{,}111$ drops, of which $56$ water.}
    \label{fig:drop_locations_area_min}
\end{figure}
\begin{figure}[htp!]
    \centering
    \includegraphics[width=0.97\textwidth]{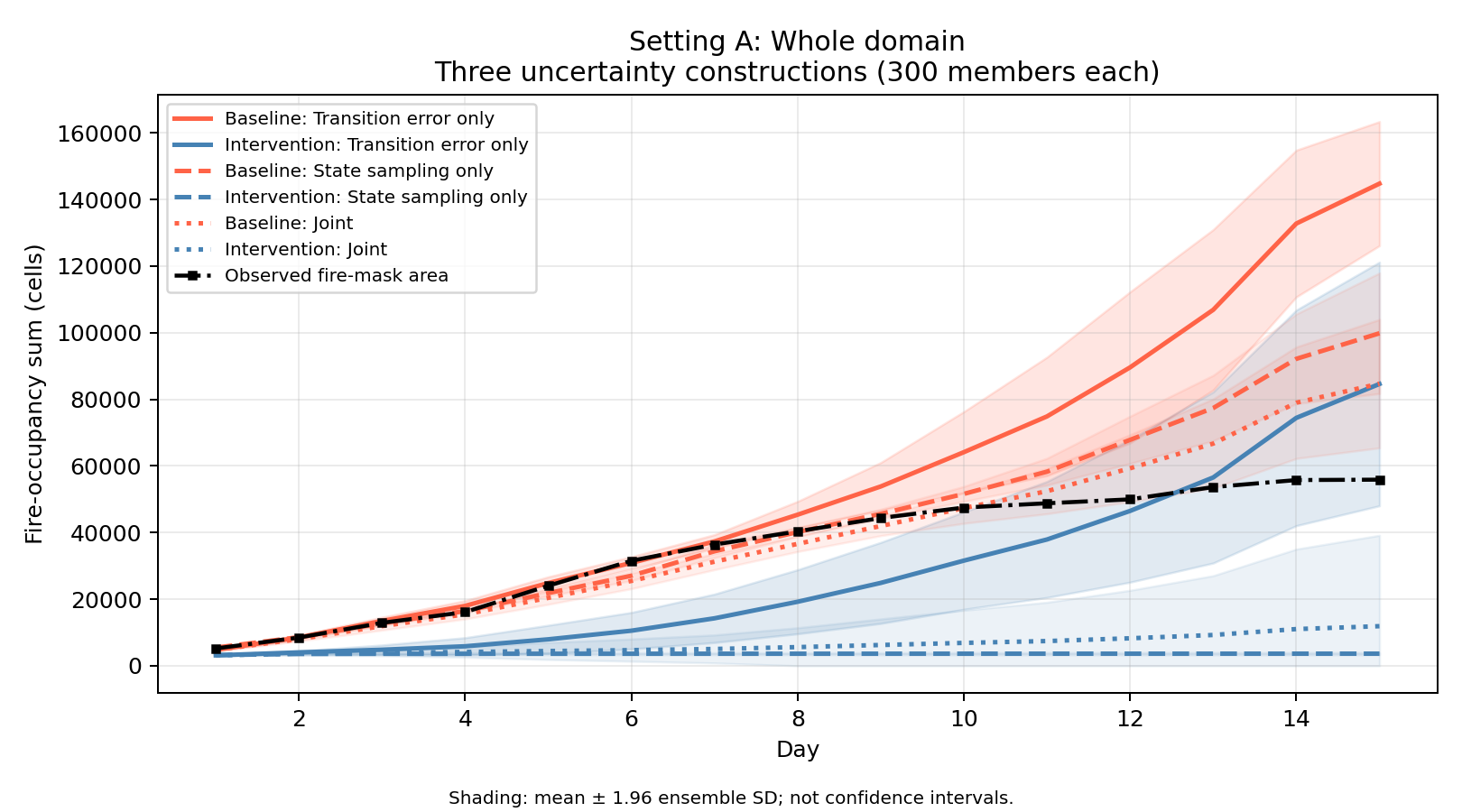}
\caption{Setting A. Daily whole-domain fire occupancy for the no-drop baseline and the $1{,}111$-drop nominal schedule, in cells ($0.09$\,ha each). Dashed curves show state sampling alone, solid curves transition error alone, and dotted curves their joint construction, with $300$ members each. The black dash-dot curve is observed fire-mask area. Shading is the mean $\pm1.96$ ensemble standard deviations, clipped below at zero; it describes spread, not a confidence interval.}
    \label{fig:total_fire_uq_area_min}
\end{figure}

Setting A's first stage reaches its best loss, $2.72$, at epoch $3{,}910$ with $2{,}727$ executed drops and is stopped $2{,}000$ epochs later without improvement, after $19$ minutes on one GPU at $0.19$\,s per epoch. The backward elimination reduces the $2{,}727$ drops to $1{,}111$ ($59\%$ fewer) in three passes, $4{,}004$ rollouts and $272$ seconds: the first pass removes $1{,}611$ drops, the second $5$, and the third none. Retardant drops decrease from $2{,}657$ to $1{,}055$ and water drops from $70$ to $56$. The retained water drops act on the burning core in the first days, where one application suppresses ignition immediately; the retained retardant lines contain the perimeter over the remaining horizon. The schedule has fire objective $2.77$ and reduces deterministic terminal extent by $89.5\%$, from $167{,}865$ to $17{,}687$ probability-cells, and time-integrated occupancy by $86.7\%$ (Figure~\ref{fig:fire_difference_map_area_min}); Figure~\ref{fig:drop_locations_area_min} shows its spatial distribution. These are reductions of mean-field probability sums. Daily state sampling gives a $96.4\%$ reduction in the Day-15 ensemble mean. Ignition-transition error without state sampling gives a $41.5\%$ reduction in mean Day-15 extent (Table~\ref{tab:uq_results_a}). Its baseline and intervention means are $13{,}027.8$ and $7{,}614.8$\,ha equivalents. The standard deviation of their paired difference is $1{,}797.2$\,ha. This transition-error-only evaluation preserves burned-state history and yields a smaller mean extent reduction than the deterministic rollout.

Figure~\ref{fig:total_fire_uq_area_min} compares the three ensemble constructions. In the joint ensemble, the baseline and intervention means are $7{,}623.5$ and $1{,}070.2$\,ha, an $86.0\%$ reduction. The paired benefit has standard deviation $1{,}498.2$\,ha. Daily categorical sampling changes both baseline and intervention trajectories through subsequent nonlinear spread. The difference from the transition-error-only result therefore measures the effect of this joint construction, not the benefit of correcting the fire state from external observations. A positive mean benefit does not imply a benefit in every member.

\begin{table}[htp!]
\centering
\footnotesize
\caption{Setting A. Day-15 whole-domain ensemble summaries in hectares for the $1{,}111$-drop nominal schedule, $n=300$ per construction. State-only and joint members give realized fire-cell counts; transition-only members give sums of $p^B+p^R$. One cell is $0.09$\,ha. Std is ensemble standard deviation, not standard error. Reductions compare means against the baseline of the same construction.}
\label{tab:uq_results_a}
\setlength{\tabcolsep}{5pt}
\begin{tabular}{l rr rr c}
\toprule
& \multicolumn{2}{c}{Baseline} & \multicolumn{2}{c}{Optimized} & Red. \\
\cmidrule(lr){2-3}\cmidrule(lr){4-5}
\textbf{Evaluation} & Mean & Std & Mean & Std & (\%) \\
\midrule
State only & 8{,}984.9 & 830.8 & 327.3 & 15.1 & 96.4 \\
Transition only & 13{,}027.8 & 855.0 & 7{,}614.8 & 1{,}679.2 & 41.5 \\
Joint & 7{,}623.5 & 885.2 & 1{,}070.2 & 1{,}252.6 & 86.0 \\
\bottomrule
\end{tabular}
\end{table}

\subsubsection{Setting B: protected-region exposure}
\label{sec:results_b}
\begin{figure}[htp!]
    \centering
    \includegraphics[width=0.97\textwidth]{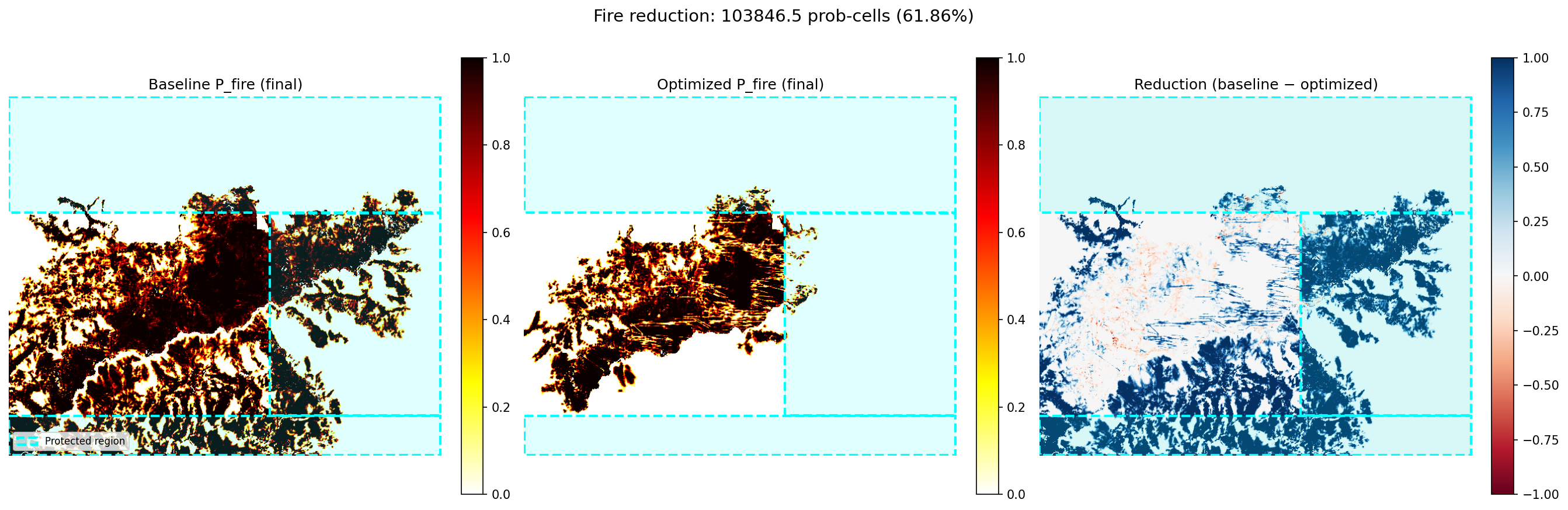}
    \caption{Setting B. Final fire-occupancy maps at the nominal effectiveness with the three protected regions outlined. The fire is contained to the unprotected central area; the deterministic rollout reduces total terminal extent by $61.9\%$ and leaves $3{,}143$ protected cells affected, $4.6\%$ of the baseline.}
    \label{fig:fire_difference_map_prot_area}
\end{figure}
\begin{figure}[htp!]
    \centering
    \includegraphics[width=0.97\textwidth]{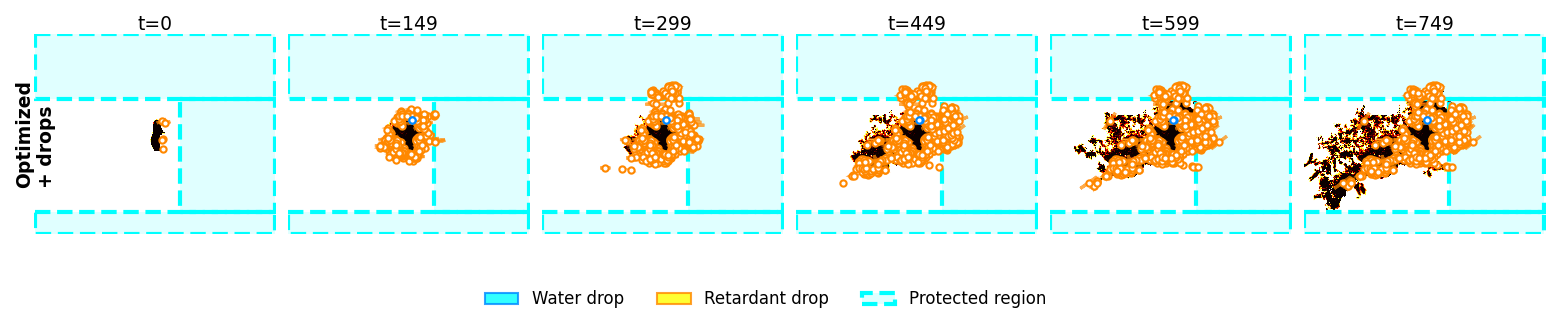}
    \caption{Setting B. Spatial distribution of the optimized drops over time, after the backward elimination: $735$ drops, two of them water.}
    \label{fig:drop_locations_prot_area}
\end{figure}
\begin{figure}[htp!]
    \centering
    \includegraphics[width=0.97\textwidth]{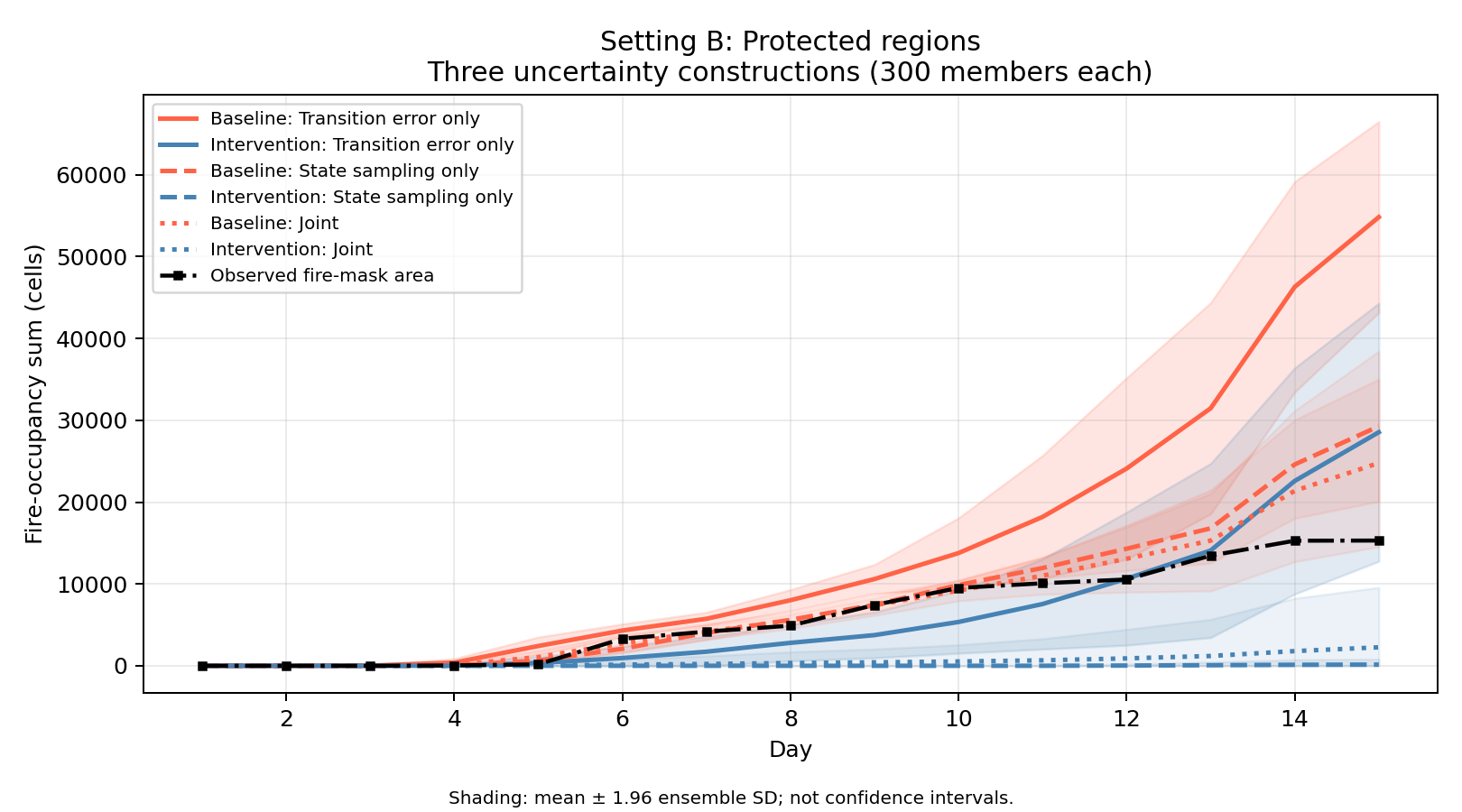}
\caption{Setting B. Daily fire occupancy summed over the protected regions for the no-drop baseline and the $735$-drop nominal schedule, in cells. Dashed curves show state sampling alone, solid curves transition error alone, and dotted curves their joint construction, with $300$ members each. The black dash-dot curve is observed fire-mask area. Shading is the mean $\pm1.96$ ensemble standard deviations, clipped below at zero; it describes spread, not a confidence interval.}
    \label{fig:prot_area_uq_prot_area}
\end{figure}

Setting B's first stage starts from Setting A's first-stage schedule, whose protected exposure is $0.0022$, above the threshold floor. It reaches its best loss, $0.201$, at epoch $3{,}970$ with $2{,}733$ executed drops and is stopped $2{,}000$ epochs later, after $19$ minutes. The backward elimination under the protection constraint alone reduces the $2{,}733$ drops to $735$ ($73\%$ fewer) in two passes, $2{,}785$ rollouts and $197$ seconds: the first pass removes $1{,}998$ drops and the second none. Water drops decrease from $70$ to $2$ and retardant drops from $2{,}663$ to $733$. The schedule leaves $3{,}143$ protected cells affected at Day~15 in the deterministic rollout, $4.6\%$ of the baseline's $67{,}909$; because neither burn nor terminal extent is constrained, the fire occupies the unprotected central area, and total terminal extent decreases by $61.9\%$ (Figure~\ref{fig:fire_difference_map_prot_area}). Figure~\ref{fig:drop_locations_prot_area} shows the schedule. As in Setting A, these are mean-field reductions. Daily state sampling gives a $99.5\%$ reduction in Day-15 protected-region exposure and $63.6\%$ in total area. Under ignition-transition error alone, the corresponding reductions are $47.9\%$ and $30.3\%$ (Table~\ref{tab:uq_results_b}). The protected-region benefit has paired standard deviation $9{,}477.4$ cells. Thus the optimized schedule retains a positive mean benefit under the fitted error model, while its deterministic protected-region reduction substantially overstates that ensemble result.

Under daily state sampling alone, the protected-region residual has mean $145.1$ cells and standard deviation $340.6$ cells. Joint state sampling and transition error give mean protected-region and whole-domain reductions of $90.9\%$ and $48.6\%$, respectively (Table~\ref{tab:uq_results_b}). These percentages use the joint ensemble's own baseline.

\begin{table}[htp!]
\centering
\footnotesize
\caption{Setting B. Day-15 summaries for the $735$-drop nominal schedule, $n=300$ per construction. Protected-region values sum occupancy over the union of the three regions, in cells; whole-domain values are hectare equivalents. Std is ensemble standard deviation. State-only and joint members give realized counts; transition-only members give probability sums. Reductions compare means with the baseline of the same construction.}
\label{tab:uq_results_b}
\setlength{\tabcolsep}{5pt}
\begin{tabular}{ll rr rr c}
\toprule
& & \multicolumn{2}{c}{Baseline} & \multicolumn{2}{c}{Optimized} & Red. \\
\cmidrule(lr){3-4}\cmidrule(lr){5-6}
\textbf{Quantity} & \textbf{Evaluation} & Mean & Std & Mean & Std & (\%) \\
\midrule
Protected regions (cells) & State only & 29{,}269.0 & 4{,}681.7 & 145.1 & 340.6 & 99.5 \\
 & Transition only & 54{,}822.6 & 5{,}953.4 & 28{,}538.9 & 8{,}045.4 & 47.9 \\
 & Joint & 24{,}781.0 & 5{,}223.4 & 2{,}257.4 & 3{,}733.9 & 90.9 \\
Total area (ha) & State only & 8{,}984.9 & 830.8 & 3{,}269.4 & 596.8 & 63.6 \\
 & Transition only & 13{,}027.8 & 855.0 & 9{,}080.6 & 1{,}287.5 & 30.3 \\
 & Joint & 7{,}623.5 & 885.2 & 3{,}921.5 & 745.5 & 48.6 \\
\bottomrule
\end{tabular}
\end{table}

\paragraph{Design effectiveness for Setting B.}
We also run both stages at design effectiveness pairs $0.85/0.75$ and $0.80/0.70$ (retardant/water), evaluating every schedule at $\eta=0.9$. Each Setting B protection solve starts from the Setting A first-stage schedule obtained at the same design effectiveness. The comparison therefore includes both the effectiveness used during optimization and the resulting warm start. For the new $0.80/0.70$ design, backward elimination reduces $3{,}061$ drops to $746$ in two passes and $2{,}746$ rollouts, while enforcing the same $2\%$ protection tolerance. Table~\ref{tab:design_eta_b} compares the three final schedules. The lower-effectiveness designs slightly improve deterministic protected-region reduction, to $96.2\%$ and $95.8\%$ versus $95.4\%$ for the nominal design. Without daily state sampling, under $300$ common ignition-error realizations, the nominal design gives the largest mean protected-region reduction: $47.9\%$, compared with $45.1\%$ and $43.0\%$ for the lower-effectiveness designs, with similar drop counts. Under joint sampling and transition error, the three protected-region reductions are $90.9\%$, $88.2\%$, and $87.7\%$, in the same design order. The nominal design therefore has the largest mean protection benefit in both error-containing ensembles for these Setting B schedules.

\begin{table}[htp!]
\centering
\footnotesize
\caption{Setting B design-effectiveness comparison. All schedules are evaluated at $\eta=0.9$. Values are Day-15 percentage reductions relative to the no-drop baseline; ``Protected'' denotes exposure inside the three designated regions and ``Total'' denotes whole-domain extent. Transition-only and joint reductions compare $300$-member means with their respective baselines. Joint adds daily categorical state sampling with the same error parameters.}
\label{tab:design_eta_b}
\setlength{\tabcolsep}{3pt}
\begin{tabular}{lrrrrrrr}
\toprule
& & \multicolumn{2}{c}{Deterministic} & \multicolumn{2}{c}{Transition only} & \multicolumn{2}{c}{Joint} \\
\cmidrule(lr){3-4}\cmidrule(lr){5-6}\cmidrule(lr){7-8}
Design $\eta^{\mathrm{d}}_{\mathrm{ret}}/\eta^{\mathrm{d}}_{\mathrm{w}}$ & Drops & Protected & Total & Protected & Total & Protected & Total \\
\midrule
0.90 / 0.90 & 735 & 95.4\% & 61.9\% & 47.9\% & 30.3\% & 90.9\% & 48.6\% \\
0.85 / 0.75 & 754 & 96.2\% & 57.2\% & 45.1\% & 25.9\% & 88.2\% & 38.2\% \\
0.80 / 0.70 & 746 & 95.8\% & 59.9\% & 43.0\% & 25.2\% & 87.7\% & 40.7\% \\
\bottomrule
\end{tabular}
\end{table}

Both objectives yield large simulator-relative reductions, but they allocate the fleet differently: Setting A limits spread throughout the domain, whereas Setting B prioritizes the designated regions.

\subsection{Planner benchmark}
\label{sec:planner_benchmark}

We compare gradient-based planning with simpler strategies under a common simulator, objective, and resource allowance.

\paragraph{Reference planners.}
The \textit{simulator baseline} applies no drops. Two \textit{random feasible} controls place drops uniformly over either the raster or the baseline burn footprint; neither matches drop times to fire arrival. A \textit{tactical heuristic} uses suppression doctrine~\citep{plucinski2019aerial}: water targets the active front, retardant targets forecast spread half a day ahead, and dispersed footprints lie across the local spread direction. It recomputes the forecast and placements for three passes and retains the best schedule. A \textit{greedy marginal-benefit} planner scores individual additions and commits improving candidates in batches, adapting the standard greedy rule~\citep{nemhauser1978greedy}. Three \textit{derivative-free} planners search activation and pose variables: CMA-ES~\citep{hansen2001cmaes,hansen2016tutorial}, differential evolution~\citep{storn1997de}, and particle swarm optimization~\citep{kennedy1995pso}. The latter two use nevergrad~\citep{rapin2018nevergrad}. Appendix~\ref{app:planners} specifies the implementations.

\paragraph{Matched conditions.}
Every schedule contains binary execution decisions and continuous poses and is evaluated through one shared objective: the setting's fire objective plus a penalty for exceeding its drop budget (\eqref{eq:shared_objective}). All planners share the fleet, grounding calendar, and turnaround gates. The budgets match the reported two-stage schedules: $1{,}111$ drops in Setting A and $735$ in Setting B. All planners, the two-stage procedure included, search and are scored at the nominal effectiveness $\eta=0.9$. The budgets are soft limits enforced through the common penalty; a schedule can exceed them and incur a cost.

Search-based planners receive $6{,}000$ rollout-equivalent evaluations. A gradient epoch includes a forward and reverse sweep and is charged as three rollouts, a conservative accounting convention; the measured cost is $2.05$ rollouts (Section~\ref{sec:scaling}), so the charge favours the other planners. The benchmark gradient planner runs only Stage~1, at the Stage~1 learning rate of $3\times10^{-3}$ on the budgeted objective, with the Stage~1 initialization of Section~\ref{sec:opt}: drop logits just below the execution threshold, positions on the front and headings across the wind. The budget penalty keeps its count at or below $B$. It is distinct from the full two-stage procedure. The derivative-free planners start from the baseline fire-front centroid and a near-budget execution schedule; the random, tactical, and greedy constructions are described separately in Appendix~\ref{app:planners}.

\paragraph{Decision dimension.}
The native space has $T N_a 4=63{,}000$ variables: one activation and three pose variables per step and aircraft. The derivative-free planners use a daily basis with $1{,}260$ variables, repeating one activation and pose through each day's steps. This reduction makes their search more tractable but restricts the schedules they can represent. Search dimensions are reported in Table~\ref{tab:planner_cost}; a native-space CMA-ES control is reported below.

\paragraph{Results.}
Tables~\ref{tab:planner_benchmark} and~\ref{tab:planner_benchmark_b} report deterministic mean-field performance; Table~\ref{tab:planner_cost} reports cost. Their fire metrics are probability sums. They must be distinguished from the state-sampling, transition-error-only, and joint ensembles in Tables~\ref{tab:uq_results_a} and~\ref{tab:uq_results_b}. All planners within a benchmark table use the same evaluation. Objective values cannot be compared across settings because the weights differ.

First, the gradient planner has the lowest mean objective at the shared allowance in both settings. In Setting A, its objective is $7.04\pm0.54$, with $72.1\%$ less final extent and $92.0\%$ less protected-region fire than the baseline. The tactical heuristic has the second-lowest objective, $15.29\pm0.61$; greedy has $17.82$, CMA-ES $18.80\pm0.68$, and differential evolution and the random controls exceed $21$. The seed ranges of the gradient planner, $6.43$ to $7.74$, and of the heuristic, $14.42$ to $15.80$, do not overlap. In Setting B the gradient objective is $1.09\pm0.28$ against $3.35$ for greedy, $3.40\pm0.19$ for the heuristic and $3.45\pm0.24$ for CMA-ES. The heuristic is substantially less expensive: seven rollouts and about five seconds, versus $6{,}002$ charged rollouts and $403$ seconds for the gradient planner. Differential evolution and particle swarm exceed the drop budget in both settings and incur the corresponding penalty.

Second, the methods respond differently to additional computation. The random and tactical implementations have fixed costs. Greedy terminates after $2{,}333$ rollouts with $1{,}103$ drops in Setting A and after $1{,}532$ rollouts at the $735$-drop budget in Setting B. The derivative-free searches stop improving, while the gradient planner continues improving over most of the allowance.

\begin{figure}[htbp]
\centering
\includegraphics[width=0.92\textwidth]{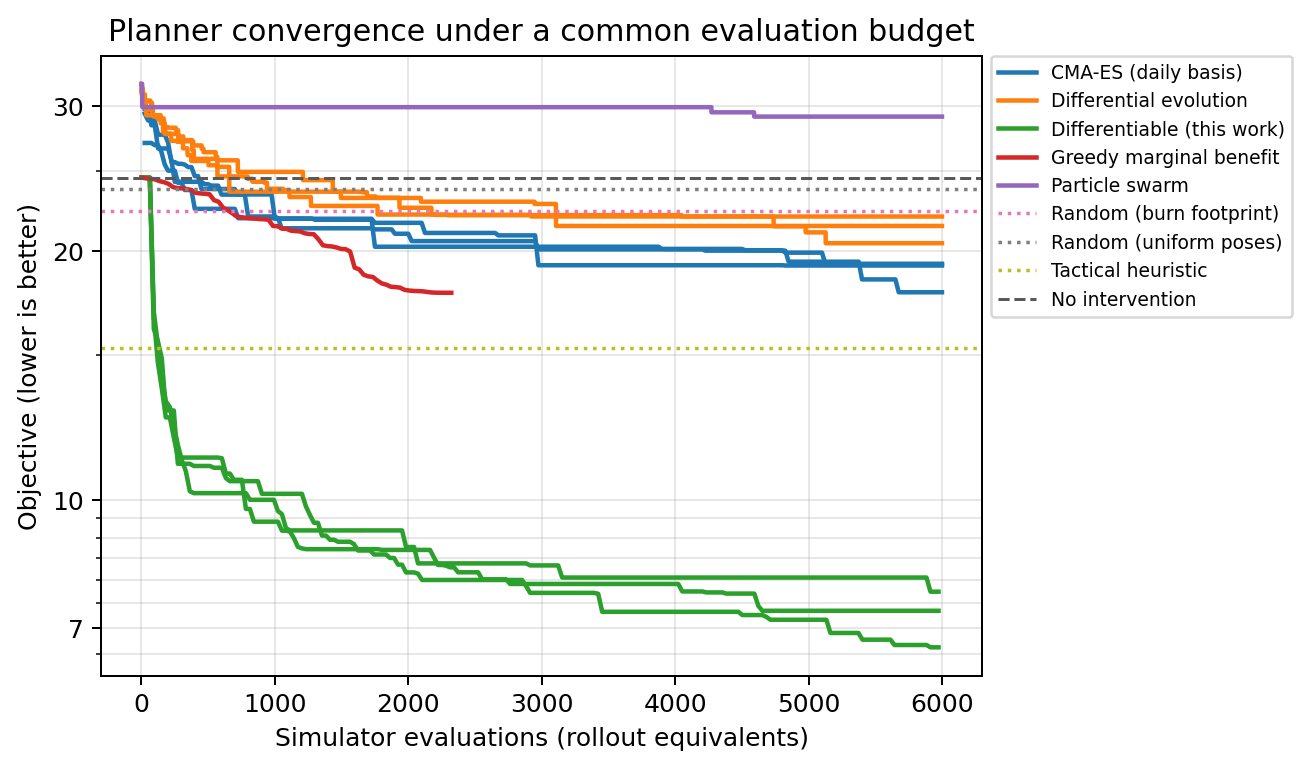}
\caption{Setting A, budget $1{,}111$ drops. Running-best objective against simulator evaluations. Solid curves are the search-based planners, one curve per seed; dotted lines mark planners that incur a fixed cost and return a single schedule, so they cannot use a larger allowance. The differentiable planner attains a lower objective than the random controls by $94$ rollouts and than the tactical heuristic by $154$ to $184$ rollouts, and its last improvements come between $4{,}650$ and $5{,}900$ rollouts. The reported two-stage schedule is lower still at $2.77$. Greedy stops after $2{,}333$ rollouts with $1{,}103$ drops. Particle swarm and differential evolution lie at or above the random controls because they exceed the drop budget and the plotted objective includes the budget penalty.}
\label{fig:planner_convergence}
\end{figure}

\begin{figure}[htbp]
\centering
\includegraphics[width=0.92\textwidth]{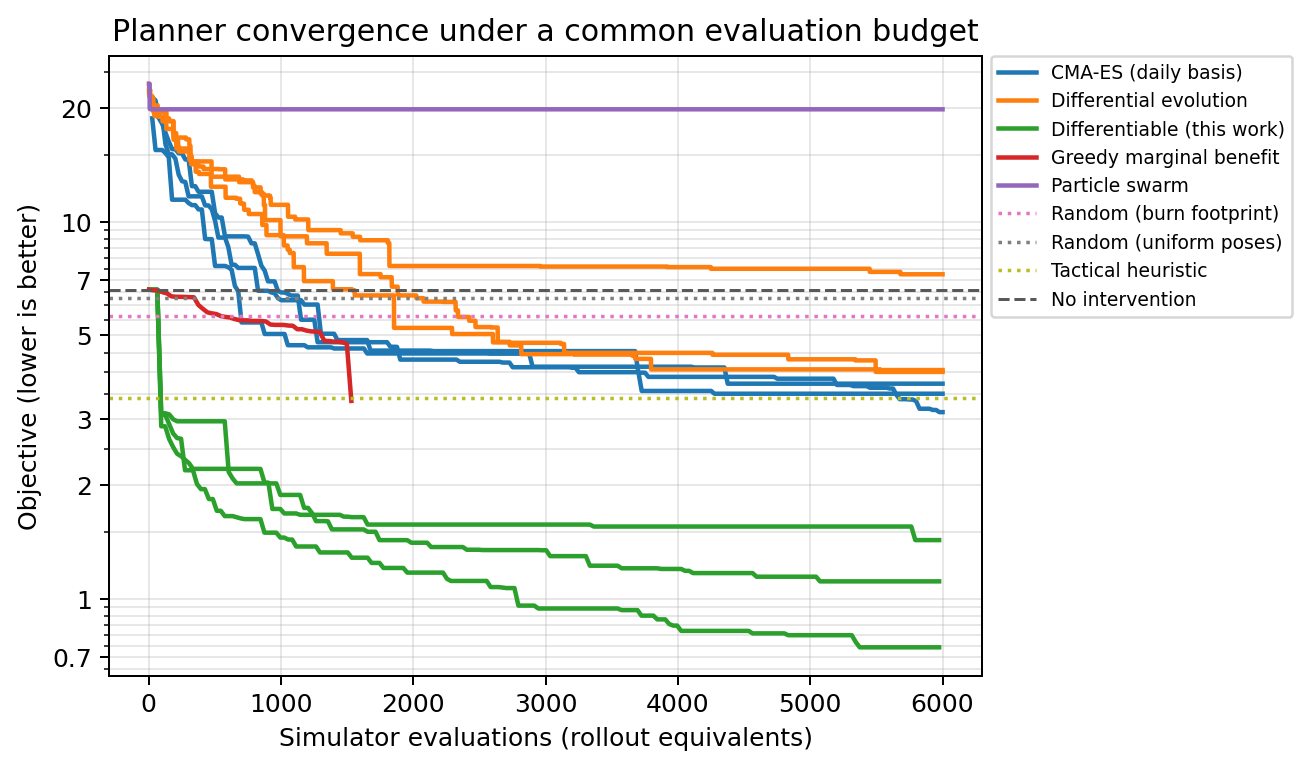}
\caption{Setting B. As Figure~\ref{fig:planner_convergence}, under the protected-region objective at a budget of $735$ drops. All three gradient-planner seeds attain a lower objective than the tactical heuristic by $94$ rollouts and end between $0.74$ and $1.43$. The reported two-stage schedule is at $0.267$ (Table~\ref{tab:planner_benchmark_b}).}
\label{fig:planner_convergence_prot}
\end{figure}

In Setting A, all three gradient seeds attain a lower objective than the random controls by $94$ rollouts and than the tactical heuristic by $154$ to $184$ rollouts, and their last improvements come between $4{,}650$ and $5{,}900$ rollouts (Figure~\ref{fig:planner_convergence}). In Setting B, all three seeds attain a lower objective than the heuristic by $94$ rollouts and improve until $5{,}100$ to $5{,}800$ rollouts (Figure~\ref{fig:planner_convergence_prot}).

Third, the full two-stage procedure achieves better performance on its selected objective. Setting A reaches objective $2.77$, with $89.5\%$ less terminal extent, $86.7\%$ less time-integrated occupancy, and $94.6\%$ less protected fire mass at Day~15. The shared-allowance gradient objective is $2.5$ times this value. In Setting B the two-stage schedule reaches objective $0.267$ and $3{,}143$ protected cells, against $1.09$ and $14{,}069\pm5{,}932$ protected cells for the gradient planner at the shared allowance, and its total extent, $64{,}018$ cells, is below the gradient planner's $94{,}914$, although Stage~2 constrained only the protected-region exposure.

The learning rate $3\times10^{-3}$ was selected in preliminary experiments with the earlier Gaussian-footprint model, using a single-seed sweep over $3\times10^{-4}$, $10^{-3}$ and $3\times10^{-3}$, with and without decay. We retained this rate for both the reported constant-flow schedules and the budgeted gradient planner. It is a shared implementation choice, not an independently established optimum for the revised footprint model.

\begin{figure}[htbp]
\centering
\includegraphics[width=0.92\textwidth]{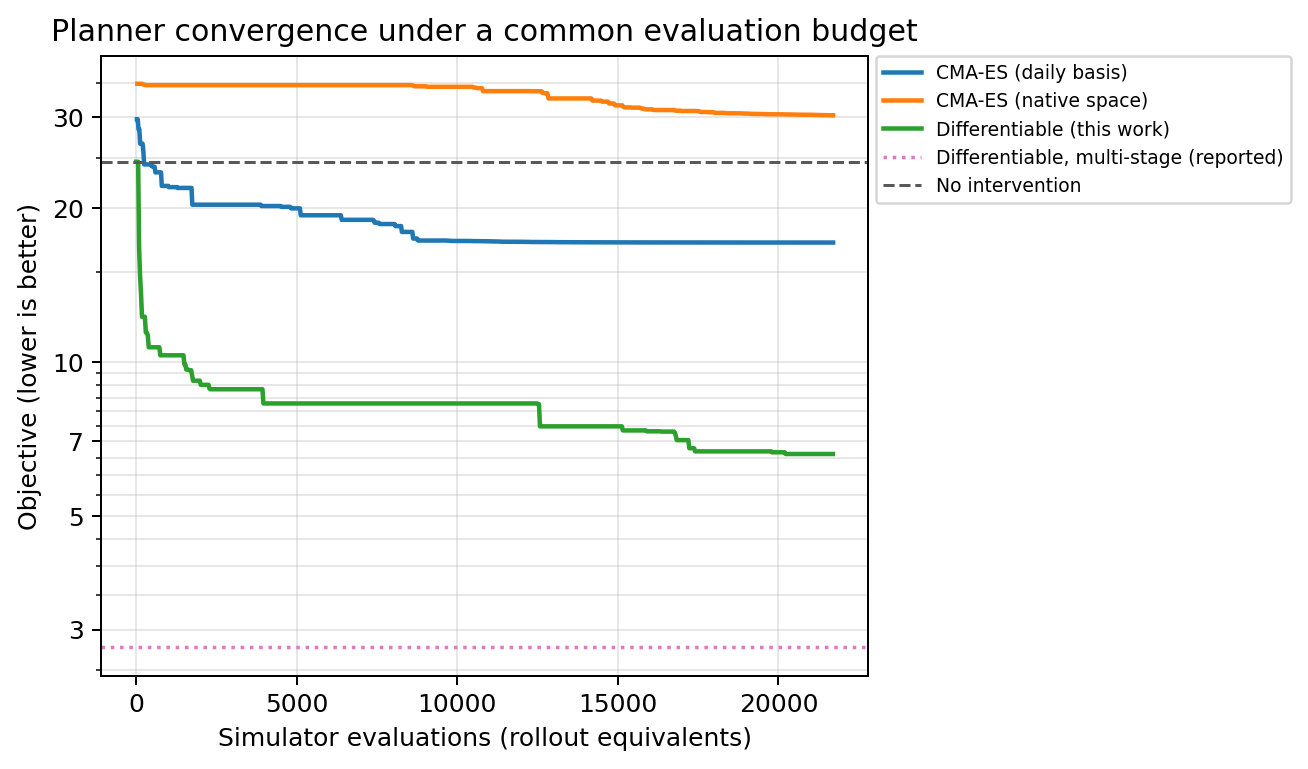}
\caption{Setting A, matched total allowance of $21{,}700$ rollouts at a budget of $1{,}111$ drops. Running-best objective for CMA-ES on the daily basis, CMA-ES in the native space, and the budgeted gradient planner; the dotted line is the reported two-stage schedule. The native-space run never attains a drop count at or below the budget, ending at $2{,}939$ drops against $1{,}111$, and the plotted objective includes the budget penalty: its fire cost alone, $13.8$, is below the no-intervention level, but the penalty for the excess drops places it above. The gradient planner last improves at $20{,}224$ rollouts and ends a factor of $2.4$ above the reported schedule.}
\label{fig:planner_convergence_matched}
\end{figure}

\paragraph{Matched total allowance.}
The two-stage procedure is not on the shared allowance: Stage~1 consumed $5{,}910$ epochs, charged as $17{,}730$ rollouts, and Stage~2 consumed $4{,}004$ rollouts, in total $21{,}734\approx21{,}700$ rollouts in Setting A; in Setting B, whose first stage ran $5{,}970$ epochs from the Setting A warm start, the total is $17{,}730+17{,}910+2{,}785\approx38{,}400$. We therefore also ran the budgeted gradient planner and, in Setting A, both CMA-ES variants at these measured costs. Table~\ref{tab:planner_matched} and Figure~\ref{fig:planner_convergence_matched} report one seed per method. In Setting A, daily-basis CMA-ES uses the whole allowance and ends at objective $17.09$ with $32.6\%$ extent reduction. Native-space diagonal CMA-ES ends with $2{,}939$ drops, substantially above the budget of $1{,}111$; its fire objective is $13.8$ before the penalty. The gradient planner reaches $6.59$ and $74.4\%$ extent reduction, with its last recorded improvement at $20{,}224$ rollouts. In Setting B the gradient planner ends at $0.925$, below the $1.09$ mean of the shared-allowance seeds.

At the matched charged allowance, the two-stage schedule has an objective lower than the budgeted gradient planner by a factor of $2.4$ in Setting A and $3.5$ in Setting B. These are single-seed comparisons of complete procedures. Setting A shares the front initialization and learning rate, while the objectives and subsequent pruning differ. In Setting B, the two-stage procedure additionally starts its protection solve from an optimized Setting A schedule, whereas the budgeted planner directly optimizes Setting B from the front initialization. Including the Setting A solve in the allowance accounts for its cost but does not isolate its benefit from that of pruning. The comparisons therefore do not identify a separate causal contribution of differentiation, warm starting, or elimination.

\begin{table}[htbp]
\centering
\caption{Matched total allowance: each planner receives the measured cost of the two-stage procedure, $21{,}700$ rollouts in Setting A and $38{,}400$ in Setting B, one seed each, at the drop budgets of $1{,}111$ and $735$. The native-space CMA-ES run never attained a drop count at or below the budget; its objective includes the budget penalty, and the value before the penalty is given in parentheses. The last row of each block is the reported two-stage schedule.}
\label{tab:planner_matched}
\footnotesize
\setlength{\tabcolsep}{3.5pt}
\begin{tabular}{lrrrrrr}
\toprule
Planner & Search dim. & Drops & Final extent & Protected & Objective & Sim. evals \\
\midrule
\multicolumn{7}{l}{\emph{Setting A, total fire-affected area}} \\
CMA-ES (daily basis) & 1,260 & 1,053 & 113,097 & 42,157 & 17.09 & 21,701 \\
CMA-ES (native space, diagonal) & 63,000 & 2,939 & 87,171 & 24,079 & 30.29 (13.8) & 21,701 \\
Budgeted gradient planner & 63,000 & 1,109 & 42,910 & 4,980 & 6.59 & 21,701 \\
Differentiable, multi-stage (reported) & 63,000 & 1,111 & 17,687 & 3,676 & 2.77 & 21,734 \\
\midrule
\multicolumn{7}{l}{\emph{Setting B, protected-region exposure}} \\
Budgeted gradient planner & 63,000 & 733 & 93,173 & 10,200 & 0.925 & 38,402 \\
Differentiable, multi-stage (reported) & 63,000 & 735 & 64,018 & 3,143 & 0.267 & 38,425 \\
\bottomrule
\end{tabular}
\end{table}

\paragraph{Schedule feasibility.}
The independent schedule audit finds zero grounded-day, turnaround, or same-aircraft simultaneity violations across the tested planners and seeds in both settings (Table~\ref{tab:schedule_feasibility}). These checks concern execution gates, not adherence to the soft drop budget.

\begin{table}[htbp]
\centering
\caption{Setting A. Suppression-planner benchmark on Bear 2020 under the frozen CNN-CA simulator and the Setting A objective (total fire-affected area), at a shared executed-drop budget of $1{,}111$ and a shared allowance of $6{,}000$ simulator rollouts (costs in Table~\ref{tab:planner_cost}). \emph{Final extent} is $\sum_x P^{\mathrm{fire}}_T(x)$ over a single deterministic rollout, \emph{Occupancy} is time-integrated fire occupancy in cell-days, and \emph{Protected} is the final fire mass inside the protected regions; all three are probability sums, and lower is better. Stochastic planners are reported as mean~$\pm$~standard deviation over three seeds. All planners search and are scored at the nominal effectiveness. The matched-allowance rows give the planners at the measured cost of the two-stage procedure, $21{,}700$ rollouts (Table~\ref{tab:planner_matched}). The multi-stage row is the reported schedule from the full two-stage procedure, which is not on the shared allowance.}
\label{tab:planner_benchmark}
\scriptsize
\setlength{\tabcolsep}{3pt}
\begin{tabular}{lrrrrr}
\toprule
Planner & Drops & Final extent & Occupancy & Protected & Objective \\
\midrule
CMA-ES (native space), matched & 2,939 & 87,171 & 613,978 & 24,079 & 30.29 \\
Particle swarm & 1,675 & 164,150 & 879,664 & 64,742 & 29.14 \\
No intervention (baseline) & 0 & 167,865 & 895,948 & 67,909 & 24.58 \\
Random (uniform poses) & 1,111 $\pm$ 0 & 162,750 $\pm$ 1,040 & 869,069 $\pm$ 4,405 & 64,904 $\pm$ 710 & 23.83 $\pm$ 0.14 \\
Random (burn footprint) & 1,111 $\pm$ 0 & 152,804 $\pm$ 3,671 & 815,723 $\pm$ 37,566 & 58,951 $\pm$ 3,886 & 22.37 $\pm$ 0.66 \\
Differential evolution & 1,255 $\pm$ 78 & 135,769 $\pm$ 842 & 746,436 $\pm$ 1,057 & 52,434 $\pm$ 777 & 21.33 $\pm$ 0.65 \\
CMA-ES (daily basis) & 1,074 $\pm$ 17 & 126,296 $\pm$ 5,045 & 716,970 $\pm$ 21,313 & 45,587 $\pm$ 3,459 & 18.80 $\pm$ 0.68 \\
Greedy marginal benefit & 1,103 & 123,864 & 617,407 & 46,689 & 17.82 \\
CMA-ES (daily basis), matched & 1,053 & 113,097 & 677,699 & 42,157 & 17.09 \\
Tactical heuristic & 1,111 & 107,972 $\pm$ 4,039 & 503,567 $\pm$ 24,700 & 38,376 $\pm$ 2,097 & 15.29 $\pm$ 0.61 \\
Differentiable (this work) & 1,110 $\pm$ 1 & 46,771 $\pm$ 3,698 & 276,205 $\pm$ 20,161 & 5,433 $\pm$ 1,337 & 7.04 $\pm$ 0.54 \\
Differentiable, matched & 1,109 & 42,910 & 272,208 & 4,980 & 6.59 \\
Differentiable, multi-stage (reported) & 1,111 & 17,687 & 119,235 & 3,676 & 2.77 \\
\bottomrule
\end{tabular}
\end{table}

\begin{table}[htbp]
\centering
\caption{Setting B. Suppression-planner benchmark on Bear 2020 under the frozen CNN-CA simulator and the Setting B objective (protected-region exposure), at a shared executed-drop budget of $735$ and a shared allowance of $6{,}000$ simulator rollouts (costs in Table~\ref{tab:planner_cost}). \emph{Final extent} is $\sum_x P^{\mathrm{fire}}_T(x)$ over a single deterministic rollout, \emph{Occupancy} is time-integrated fire occupancy in cell-days, and \emph{Protected} is the final fire mass inside the protected regions; all three are probability sums, and lower is better. Stochastic planners are reported as mean~$\pm$~standard deviation over three seeds. All planners search and are scored at the nominal effectiveness. The matched-allowance rows gives the gradient planner at the measured cost of the two-stage procedure, $38{,}400$ rollouts (Table~\ref{tab:planner_matched}). The multi-stage row is the reported schedule from the full two-stage procedure, which is not on the shared allowance.}
\label{tab:planner_benchmark_b}
\scriptsize
\setlength{\tabcolsep}{3pt}
\begin{tabular}{lrrrrr}
\toprule
Planner & Drops & Final extent & Occupancy & Protected & Objective \\
\midrule
Particle swarm & 1,713 & 167,315 & 892,801 & 67,678 & 19.88 \\
No intervention (baseline) & 0 & 167,865 & 895,948 & 67,909 & 6.61 \\
Random (uniform poses) & 735 $\pm$ 1 & 163,452 $\pm$ 1,204 & 874,695 $\pm$ 6,960 & 64,992 $\pm$ 896 & 6.28 $\pm$ 0.10 \\
Random (burn footprint) & 735 $\pm$ 1 & 157,943 $\pm$ 2,115 & 835,671 $\pm$ 34,587 & 62,347 $\pm$ 2,220 & 5.62 $\pm$ 0.78 \\
Differential evolution & 781 $\pm$ 90 & 145,170 $\pm$ 11,717 & 775,200 $\pm$ 28,942 & 53,717 $\pm$ 6,789 & 5.10 $\pm$ 1.53 \\
CMA-ES (daily basis) & 702 $\pm$ 10 & 123,328 $\pm$ 2,943 & 722,028 $\pm$ 14,507 & 38,239 $\pm$ 3,335 & 3.45 $\pm$ 0.24 \\
Tactical heuristic & 735 & 125,714 $\pm$ 3,638 & 556,460 $\pm$ 26,365 & 46,701 $\pm$ 1,811 & 3.40 $\pm$ 0.19 \\
Greedy marginal benefit & 735 & 117,956 & 625,090 & 41,669 & 3.35 \\
Differentiable (this work) & 726 $\pm$ 4 & 94,914 $\pm$ 12,445 & 554,079 $\pm$ 30,309 & 14,069 $\pm$ 5,932 & 1.09 $\pm$ 0.28 \\
Differentiable, matched & 733 & 93,173 & 573,433 & 10,200 & 0.925 \\
Differentiable, multi-stage (reported) & 735 & 64,018 & 349,482 & 3,143 & 0.267 \\
\bottomrule
\end{tabular}
\end{table}

\begin{table}[htbp]
\centering
\caption{Setting A. Search dimension and cost of each planner under the shared allowance of $6{,}000$ simulator rollouts; the rows marked matched allowance are the runs of Table~\ref{tab:planner_matched} at the measured cost of the two-stage procedure, $21{,}700$ rollouts. One gradient epoch is a forward rollout plus a reverse sweep and is charged as three rollouts. \emph{Search dim.} is the number of decision variables the planner actually searches: the derivative-free methods use a reduced daily pose basis, because the native $63{,}000$-dimensional space is beyond their reach. The tactical heuristic and the random planners are fixed-cost and cannot use a larger allowance; greedy stops when the drop budget is exhausted rather than when its allowance is. The multi-stage row is the reported solution from the full two-stage procedure, which is not on the shared allowance; its columns give the wall clock of both stages and the rollouts they consumed, a Stage~1 epoch charged as three rollouts and Stage~2 as the rollouts of its backward elimination.}
\label{tab:planner_cost}
\begin{tabular}{lrrr}
\toprule
Planner & Search dim. & Sim. evals & Runtime (s) \\
\midrule
CMA-ES (native space), matched & 63,000 & 21,701 & 1,148.1 \\
Particle swarm & 1,260 & 6,001 & 290.9 \\
No intervention (baseline) & 0 & 1 & $<$0.01 \\
Random (uniform poses) & 0 & 1 & 0.03 $\pm$ 0.0047 \\
Random (burn footprint) & 0 & 1 & 0.11 $\pm$ 0.0047 \\
Differential evolution & 1,260 & 6,001 & 296.9 $\pm$ 0.3 \\
CMA-ES (daily basis) & 1,260 & 6,001 & 289.0 $\pm$ 0.79 \\
Greedy marginal benefit & 0 & 2,333 & 180.1 \\
CMA-ES (daily basis), matched & 1,260 & 21,701 & 1,055.6 \\
Tactical heuristic & 0 & 7 & 4.77 $\pm$ 0.081 \\
Differentiable (this work) & 63,000 & 6,002 & 402.9 $\pm$ 0.81 \\
Differentiable, matched & 63,000 & 21,701 & 1,403.1 \\
Differentiable, multi-stage (reported) & 63,000 & 21,734 & 1,391.0 \\
\bottomrule
\end{tabular}
\end{table}

\begin{table}[htbp]
\centering
\caption{Setting A. Schedule-feasibility audit, first seed of each planner. Candidate drops are proposals before gating; executed drops are those the simulator actually performs. The three violation counts are re-derived independently from each schedule rather than read back from simulator state, so the audit can detect a planner that emits an infeasible plan: a grounded-day violation is a drop on a day that aircraft is grounded, a turnaround violation is a pair of drops by one aircraft closer together than its turnaround time, and a simultaneity violation is more than one drop by one aircraft in one step. Every planner records zero violations of all three kinds; the audit of the Setting B schedules, not shown, also records zero.}
\label{tab:schedule_feasibility}
\footnotesize
\setlength{\tabcolsep}{3pt}
\begin{tabular}{lrrrrr}
\toprule
Planner & Candidate & Executed & Grounded & Turnaround & Simultaneous \\
\midrule
Random (uniform poses) & 1,630 & 1,111 & 0 & 0 & 0 \\
Tactical heuristic & 1,111 & 1,111 & 0 & 0 & 0 \\
Greedy marginal benefit & 1,103 & 1,103 & 0 & 0 & 0 \\
Random (burn footprint) & 1,630 & 1,111 & 0 & 0 & 0 \\
CMA-ES (daily basis) & 5,050 & 1,080 & 0 & 0 & 0 \\
Differentiable, multi-stage (reported) & 1,111 & 1,111 & 0 & 0 & 0 \\
Differential evolution & 5,200 & 1,156 & 0 & 0 & 0 \\
Differentiable (this work) & 2,737 & 1,108 & 0 & 0 & 0 \\
Particle swarm & 6,700 & 1,675 & 0 & 0 & 0 \\
No intervention (baseline) & 0 & 0 & 0 & 0 & 0 \\
Differentiable, matched & 3,070 & 1,109 & 0 & 0 & 0 \\
CMA-ES (daily basis), matched & 4,950 & 1,053 & 0 & 0 & 0 \\
CMA-ES (native space), matched & 7,680 & 2,939 & 0 & 0 & 0 \\
\bottomrule
\end{tabular}
\end{table}

\subsection{Computational scaling}
\label{sec:scaling}

We measure scaling across fifteen combinations of planning horizon, fleet size, and raster resolution. Decision dimensions range from $2{,}400$ to $63{,}000$ and raster sizes from $28{,}798$ to $463{,}012$ cells. For each setting, we time a forward rollout and an Adam step that computes derivatives for all decisions.

Table~\ref{tab:scalability} reports medians and interquartile ranges from fifteen interleaved repeats after warm-up. Interleaving reduces timing drift; repeated measurements account for allocator noise from the forward rollout's approximately $1.4$\,GB history array. Across settings, an Adam step costs $2.05\pm0.18$ forward rollouts, with a range of $1.71$ to $2.47$. The log-log slope of this ratio against decision dimension is $-0.026$, showing no increasing relative overhead in the tested range. In the fixed-raster, fixed-horizon fleet sweep, increasing $d$ from $2{,}400$ to $25{,}200$ changes the ratio from $2.24$ to $1.93$. Absolute rollout cost still increases with fleet size. This approximately constant relative overhead is consistent with reverse-mode differentiation~\citep{griewank2008evaluating}.

A coordinate-wise forward-difference gradient in $63{,}000$ variables would require $63{,}001$ rollouts. The reverse sweep provides derivatives through the implemented surrogate computation at about two rollouts per update. Derivative-free methods avoid derivatives but incur increasing search cost with dimension~\citep{kolda2003direct}, motivating the reduced basis used in the benchmark.

\begin{table}[htbp]
\centering
\caption{Cost of a full gradient relative to one forward rollout, across problem sizes. A rollout is one forward simulation; a gradient step is one Adam update returning derivatives with respect to all $d$ decision variables. Each entry is the median of 15 interleaved repeats after warm-up, with the interquartile range as a percentage of the median.}
\label{tab:scalability}
\footnotesize
\setlength{\tabcolsep}{4pt}
\begin{tabular}{lrrrrrr}
\toprule
Setting & Grid & $T$ & $N_a$ & $d$ & Rollout (ms) & Grad./rollout \\
\midrule
fleet: 2 aircraft & 619$\times$748 & 300 & 2 & 2,400 & 31.6\,(1\%) & 2.24$\times$ \\
fleet: 5 aircraft & 619$\times$748 & 300 & 5 & 6,000 & 33.0\,(2\%) & 2.23$\times$ \\
fleet: 10 aircraft & 619$\times$748 & 300 & 10 & 12,000 & 35.9\,(1\%) & 2.03$\times$ \\
fleet: 15 aircraft & 619$\times$748 & 300 & 15 & 18,000 & 37.9\,(1\%) & 1.94$\times$ \\
fleet: 21 aircraft & 619$\times$748 & 300 & 21 & 25,200 & 41.7\,(1\%) & 1.93$\times$ \\
grid: 154x187 & 154$\times$187 & 300 & 21 & 25,200 & 13.9\,(17\%) & 2.47$\times$ \\
grid: 247x299 & 247$\times$299 & 300 & 21 & 25,200 & 15.4\,(3\%) & 2.24$\times$ \\
grid: 371x448 & 371$\times$448 & 300 & 21 & 25,200 & 19.3\,(2\%) & 2.10$\times$ \\
grid: 464x561 & 464$\times$561 & 300 & 21 & 25,200 & 39.1\,(1\%) & 2.01$\times$ \\
grid: 619x748 & 619$\times$748 & 300 & 21 & 25,200 & 41.7\,(1\%) & 1.96$\times$ \\
horizon: 3d & 619$\times$748 & 150 & 21 & 12,600 & 23.5\,(4\%) & 1.71$\times$ \\
horizon: 6d & 619$\times$748 & 300 & 21 & 25,200 & 42.0\,(1\%) & 1.92$\times$ \\
horizon: 9d & 619$\times$748 & 450 & 21 & 37,800 & 60.5\,(1\%) & 2.00$\times$ \\
horizon: 12d & 619$\times$748 & 600 & 21 & 50,400 & 81.0\,(1\%) & 1.99$\times$ \\
horizon: 15d & 619$\times$748 & 750 & 21 & 63,000 & 100.8\,(0\%) & 2.00$\times$ \\
\bottomrule
\end{tabular}
\end{table}

To assess solution quality as dimension grows, we compare the gradient planner and CMA-ES across fleet sizes with the raster and horizon fixed. Each receives $600$ rollout-equivalent evaluations, with three runs per size. Fleet size changes both the decision dimension and available suppression capacity; rollout cost need not remain identical.

\begin{figure}[htbp]
\centering
\includegraphics[width=0.98\textwidth]{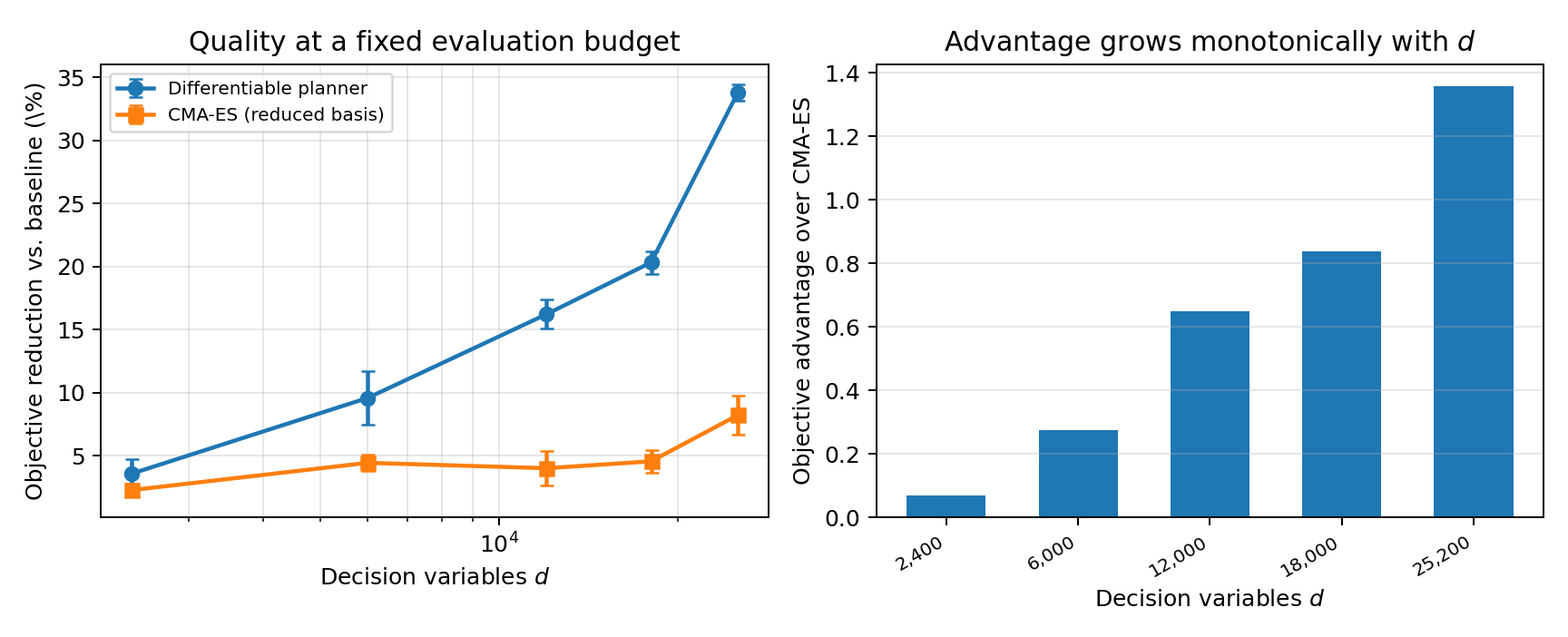}
\caption{Solution quality versus native decision dimension at $600$ rollout-equivalent evaluations, three runs per point, with one-standard-deviation error bars. Fleet size varies while raster and horizon remain fixed, changing both dimension and suppression capacity. The gradient planner uses the front initialization of Section~\ref{sec:opt}. Left: fraction of baseline objective removed. Right: gradient planner's objective advantage over CMA-ES.}
\label{fig:planner_dimension}
\end{figure}

The gradient planner attains the lower objective at every tested size (Figure~\ref{fig:planner_dimension}). At $d=2{,}400$, it reduces the baseline objective by $3.6\%$, versus $2.3\%$ for CMA-ES. At $d=25{,}200$, the reductions are $33.8\%$ and $8.2\%$. The objective gap grows monotonically from $0.07$ to $1.36$, with disjoint one-standard-deviation intervals at the four larger sizes. These results show that gradient search exploits the larger fleet more effectively under this evaluation allowance.

The timing sweep reaches the paper's full $63{,}000$-variable configuration; the solution-quality sweep reaches $25{,}200$. Within these ranges, adding decisions does not increase the gradient's cost relative to a forward rollout, and the gradient planner's quality advantage grows with fleet size.

The measured overhead is specific to this implementation and GPU. It does not bound the number of optimization steps: the gradient planner first beats the tactical heuristic after roughly one hundred to two hundred charged rollouts in the reported benchmarks, and further improvements require thousands. Moreover, varying the fleet changes the planning problem as well as its dimension. A sweep that added variables while preserving the underlying problem would isolate dimensional effects more directly.

\section{Discussion and Comparison with State of the Art}
\label{sec:discussion}
\label{sec:comparison}

We relate the method to prior work, interpret its computational trade-offs, and state the limits of the evidence.

\subsection{Relation to prior work}

Our contribution links individual aerial-drop decisions to later fire outcomes through a differentiable simulator. Operational and semi-empirical models support forward fire and treatment analyses~\citep{finney1998farsite,finney2002mtt,finney2006flammap}; differentiable models also support calibration and learning environments~\citep{xia2025pytorchfire,cakir2025jaxwildfire,maksym2026}. Here, gradients guide drop location and orientation as well as binary execution through a surrogate estimator.

Resource-allocation and initial-attack models address assignments, dispatch, routing, and containment~\citep{ntaimo2013initialattack,avci2024wildfire,granda2023decision}. Spatial suppression optimization also couples intervention placement to changing fire behavior: \citet{belval2015mixed} jointly model suppression locations, spread paths, arrival times, and fireline intensities. Our distinction is the optimization of individual aerial-drop execution, location, and orientation through learned differentiable spatial transitions, with separate material effects and aircraft constraints.

The common-simulator benchmark in Section~\ref{sec:planner_benchmark} compares adapted planning strategies for this formulation. All produce event-specific open-loop schedules.

\subsection{Planner selection by decision type}
\label{sec:which_planner}

The benchmark reveals a trade-off between computational cost and solution quality. The tactical heuristic is economical at small allowances; gradient search achieves better objectives as the allowance grows.

\paragraph{The trade-off.}
In Setting A, the tactical heuristic reduces terminal extent by $35.7\%$ on average in about five seconds. The gradient planner reaches $72.1\%$ at the shared allowance in about seven minutes and $74.4\%$ at the matched allowance in about $23$ minutes. The full two-stage procedure reaches $89.5\%$ in about the same time. Daily-basis CMA-ES remains between $25\%$ and $33\%$ at both allowances. The observed advantage of gradient search is its continued improvement with computation, while the fixed-cost heuristic provides an inexpensive starting point.

\paragraph{Cost measured against operational tempo.}
Planning times must be assessed against the decision deadline. In-flight retasking allows minutes; an operational-period plan typically allows $12$ to $24$ hours; fleet and contingency studies allow longer. Table~\ref{tab:planner_deployment} maps the measured runtimes to these potential uses, conditional on operational validation. Setting A's best first-stage parameters occur at epoch $3{,}910$, about $12$ minutes at $0.19$ seconds per epoch. Continuing for $2{,}000$ epochs without improvement takes the first stage to about $19$ minutes. The backward elimination adds about four and a half minutes over $4{,}004$ rollouts. This computation of about $24$ minutes is within an operational planning period by a wide margin, while the tactical heuristic is better suited to in-flight retasking.

\begin{table}[htbp]
\centering
\caption{Planners assigned to the decision tempos of aerial suppression. Runtimes are those of Table~\ref{tab:planner_cost} on one GPU; extent reductions are deterministic Setting A results against the simulator baseline. The mapping is conditional on a simulator that has been validated for counterfactual suppression response, which the present work does not establish.}
\label{tab:planner_deployment}
\footnotesize
\setlength{\tabcolsep}{4pt}
\begin{tabular}{lllr}
\toprule
Decision & Time available & Suitable planner & Extent reduction \\
\midrule
In-flight retasking & minutes & Tactical heuristic (5\,s) & $35.7\%$ \\
Re-planning within a period & tens of minutes & Budgeted gradient planner (7\,min) & $72.1\%$ \\
Operational-period plan & $12$ to $24$\,h & Differentiable, two-stage (about 24\,min) & $89.5\%$ \\
Pre-positioning, fleet studies & days & Differentiable, many scenarios & --- \\
\bottomrule
\end{tabular}
\end{table}

\paragraph{Complementary rather than competing use.}
The methods suggest complementary uses. A tactical schedule could initialize gradient search; optimized schedules could reveal placement patterns for faster rules; and repeated gradient optimizations could support fleet or grounding-calendar sensitivity studies. These are proposed extensions, not combinations tested here.

\paragraph{Limits of the comparison.}
The two-stage procedure received more development and tuning than the reference planners. The gradient planner also uses dispersed front initialization, whereas the derivative-free planners start near daily front centroids and generally use a reduced decision basis. The native-space CMA-ES control changes the basis but retains its centroid initialization. The results rank these implementations under the stated allowances; they do not isolate differentiation from initialization, representation, or tuning. Multi-start tactical search or continuous refinement of greedy schedules could change the comparison. All rankings are conditional on the simulator, whose response errors may be exploited by extensive optimization. The experiments do not establish which planner is most robust to such errors.

\paragraph{One solve of a receding-horizon scheme.}
A receding-horizon extension would repeatedly observe the fire, optimize the remaining horizon, and execute only the first day's decisions~\citep{garcia1989mpc,rawlings2017mpc,qin2003mpc}. This paper supplies a single planning optimization, not that feedback loop. A shifted previous schedule could provide a warm start after updating the aircraft state and feasibility constraints. A state estimator mapping perimeter observations to the three-state field is still needed.

\subsection{Limitations and practical consequences}
\label{sec:limitations}

\paragraph{Scope of the wildfire model.}
The model architecture and predictive evaluation belong to \citet{maksym2026}. The present constant-flow experiments report multi-seed planner variability and schedule sensitivity to effectiveness and pose perturbations. Preliminary initialization, budget-weight, and learning-rate studies used the earlier Gaussian-footprint model. Corresponding controlled ablations for the revised deposition model, including the straight-through threshold and logit clamp, remain untested.

\paragraph{Limitation 1: Planner comparison and scale.}
The comparison establishes relative performance within one simulator and event. It omits two families of planner. One is exact mixed-integer optimization. The problem is a mixed-integer nonlinear program, with binary drop decisions, continuous poses, and nonlinear fire dynamics, and global solvers for that class do not scale to a $750$-step rollout on a raster of this size. The practical alternative is a mixed-integer linear surrogate, obtained by linearizing or tabulating the spread model on a coarse grid and a short horizon in the manner of~\citet{belval2015mixed}. Such a solver would certify optimality for the surrogate, not for the model we plan through, so it would bound the achievable objective only up to the linearization error. The other is a reinforcement-learning policy, a network trained over many simulated episodes to map the observed fire state to drop actions, of the kind the differentiable simulators of~\citep{xia2025pytorchfire,cakir2025jaxwildfire} are built to train; our method is not of that kind, since it learns no policy and solves one optimization problem per fire. The two differ in how they are exposed to model error. A policy trained by reinforcement learning is fitted to the simulator during training, so the simulator's errors enter every decision the policy makes afterwards, and the degradation of such a policy with model inaccuracy is not known for this problem. A receding-horizon scheme built on our solver would also plan through the model, while updating its initial state from observations before each solve. This can correct state discrepancies but does not remove errors in the transition or suppression model. An observation-driven reinforcement-learning policy could likewise respond to the current fire state. Neither approach has been evaluated here under feedback, so their relative robustness remains open. The daily basis restricts the derivative-free planners' reachable schedules, while the fleet sweep changes both dimension and suppression capacity. Finally, three seeds provide only a coarse ranking where performance spreads overlap. More repetitions and comparisons in a common native space would strengthen the evidence.

\paragraph{Limitation 2: Counterfactual suppression response.}
Predictive accuracy does not validate counterfactual water or retardant effects, or their gradients. The model's predictive evidence (Section~\ref{sec:provenance}) therefore does not establish operational validity of the intervention layer. Suppression-effectiveness calibration, perturbation tests, and pose-gradient checks remain necessary.

\paragraph{Limitation 3: Intervention-event and fleet coverage.}
The intervention case study covers Bear Fire and one assumed fleet. It does not establish robustness across fires, and the schedule itself is not transferable: it is a solution for one fire, one fleet, and one weather sequence. The intervention model, optimization procedure, and uncertainty-analysis method can be applied to other events. Doing so requires suitable fire-model and environmental inputs and recalibration of the ignition-transition error parameters on the new event; the fitted Bear error parameters are not assumed transferable. That transfer has not been performed, so broader evaluation must vary events, fleet composition, weather, terrain, fuel, ignition geometry, intervention timing, and material effectiveness.

\paragraph{Limitation 4: Meteorological information.}
The planner uses the historical ERA5-Land wind sequence throughout the horizon, providing an optimistic reference based on realized weather. Forecast forcing could replace that sequence, but we have not measured degradation under forecast error or the value of updated forecasts and re-planning. These require forecast-ensemble intervention studies.

\paragraph{Limitation 5: Operational detail.}
The model includes grounded days and per-aircraft turnaround times; the turnaround times, between $0.18$ and $2.2$\,h (Table~\ref{tab:aircraft_params}), are our estimates of the transit and reload cycle of each type from an assumed base, so the bases enter only through them. The gate adds a full step beyond rounded-up turnaround, giving effective cycles of $0.96$ to $2.88$\,h at the reported resolution. Capacity and comparisons are conditional on that conservative discretization. The model omits explicit base locations and routing, airspace deconfliction, conflicts between aircraft, detailed replenishment, dynamic dispatch, and local flight-safety constraints. The fleet and its grounded days are hypothetical. Footprints are uniform constant-flow segments convolved with a Gaussian dispersion kernel, calibrated to the drop-guide tables for three aircraft and transferred by analogue to the others (Appendix~\ref{app:Intervention Optimization}); the Gaussian form of the kernel and the level-3 flow setting are modelling choices, and wind drift remains stylized. The suppression effectiveness $\eta$ is an assumption informed by estimates for Australian fuels~\citep{loane1986aerial}; it is neither measured nor calibrated on this event, the estimates that inform it concern fires below about $1{,}700$\,kW/m, and it is applied to water although the same source implies a lower water value. These omissions prevent direct operational use.

\paragraph{Limitation 6: Conditional uncertainty.}
All three ensemble constructions condition on the frozen simulator and fixed schedule. Daily state sampling explores the stipulated categorical process. The ignition-transition error model preserves the allowed state transfers and is fitted through rollouts on Days~1--10. On Days~11--15, the $300$-member baseline has spatial energy score $0.2125$, compared with $0.3373$ for the deterministic rollout. This score improvement does not remove the late overprediction: the observed Day-15 fire mask covers $5{,}029.2$\,ha, compared with the transition-error baseline mean of $13{,}027.8$\,ha. This is a temporal check on Bear, not validation on an independent fire. The selected correlation length is the shortest tested in the final grid; its value should not be interpreted as an independently resolved physical error scale. Spatial stationarity, daily independence, and fixed burn duration remain modeling assumptions. Observed discrepancies also include sensing error and historical suppression. The joint experiment retains the parameters fitted without daily categorical resets; its results characterize that specified combination. Daily sampling changes the stochastic propagation and does not measure the value of external observations or feedback control. Paired evaluation quantifies conditional differences, but cannot establish historical acreage saved.

\paragraph{Elimination against a deterministic metric.}
The backward elimination preserves selected deterministic metrics at the inputs used during design. It does not guarantee tolerance to placement or model errors. In Setting A, the nominal schedule's objective is multiplied by $1.7$ on average under the $10^{-4}$ pose-logit perturbations of Section~\ref{sec:design_eta}. Designing at $0.85/0.75$ reduces this pose sensitivity and improves the transition-error-only reduction from $41.5\%$ to $60.5\%$, at the cost of $67\%$ more drops. It does not improve the deterministic or state-sampling reduction. In the joint ensemble, that Setting A design increases the reduction from $86.0\%$ to $90.8\%$, while the $0.80/0.70$ design gives $76.7\%$. In Setting B, the nominal design gives the largest mean protected-region reduction under both transition error alone and the joint construction. Effectiveness sensitivity is also present before pruning. These results motivate robustness-aware design and evaluation; they do not establish that feedback re-optimization is the only remedy or that it would recover the nominal performance.

\section{Conclusion}

We presented a two-stage aerial-intervention framework that optimizes binary execution and continuous drop geometry through a frozen CNN-CA simulator, then removes drops by backward elimination subject to evaluated fire-performance tolerances. Both stages of the nominal schedules use the nominal suppression effectiveness. In Setting A, the $0.85/0.75$ design reduces sensitivity to the tested small pose perturbations and improves the mean extent reduction under both transition error alone and joint state sampling and transition error, at the cost of more drops and a worse deterministic objective. In Setting B, the nominal design has the largest mean protected-region reduction among the three tested designs under both error-containing ensembles. The Bear Fire study demonstrates total-area and protected-region objectives and evaluates fixed schedules under daily state sampling, fitted ignition-transition error, and their joint construction.

The tactical heuristic provides a substantial extent reduction in seconds. The tested gradient implementation improves on it with larger evaluation allowances, and the full two-stage Setting A schedule reduces deterministic terminal extent by $89.5\%$ relative to the simulator baseline with $1{,}111$ drops. Gradient computation costs about two forward rollouts across the tested problem sizes. These results support computational feasibility and solution quality for the tested implementations and simulator. Without daily state sampling, the fitted transition-error evaluation preserves the allowed state transfers and gives mean reductions of $41.5\%$ in Setting A terminal extent and $47.9\%$ in Setting B protected-region exposure. With joint state sampling and transition error, the nominal reductions are $86.0\%$ in Setting A whole-domain extent and $90.9\%$ in Setting B protected-region exposure. The differences between ensemble constructions, together with pose and effectiveness sensitivity, motivate evaluating designs beyond their nominal fire objective. The results do not establish historical acreage saved or operational validity.

The main next steps are calibrated suppression response, multi-event intervention evaluation, and robustness-aware pruning. A receding-horizon extension could update schedules as fire observations and weather forecasts change.


\begin{appendices}

\section{Intervention Optimization}
\label{app:Intervention Optimization}

This appendix lists the experimental parameters and describes drop-footprint calibration and the assumed suppression response.

\subsection{Optimization setup and parameters}
\label{app:Optimization setup and parameters}

\begin{table*}[htp!]
\centering
\footnotesize
\caption{Main optimization and rollout parameters, as archived with the reported solution. Initialization entries describe Setting A; Setting B starts from its optimized first-stage schedule. The epoch count is an allowance; the reported runs stopped earlier as described below.}
\label{tab:opt_params}
\begin{tabular}{p{3.6cm}p{1cm}p{6.4cm}}
\toprule
\textbf{Parameter} & \textbf{Value} & \textbf{Description} \\
\midrule
Micro-steps per day $S_{\mathrm{day}}$ & 50 & Temporal resolution of the CA rollout. \\
Grid spacing $\Delta x$ & 30.0 m & Raster resolution. \\
Epochs & 10{,}000 & Cap on the first stage. Setting A attained its best objective at epoch $3{,}910$ and terminated by its patience rule at $5{,}910$; Setting B, initialized from that schedule, attained its best at epoch $3{,}970$ and terminated at $5{,}970$; the second stage terminates by its own rule (Section~\ref{sec:step5}). \\
Learning rate & $3\times10^{-3}$ & First stage, both settings; the second stage takes no gradient steps. \\
Patience & 2{,}000 & Epochs without improvement after which the first stage stops. \\
Gradient clip norm & 1.0 & Global gradient clipping threshold. \\
Drop-logit clamp & 3.0 & Post-update clamp for drop logits. \\
Initial drop logit & $-0.04$ & Every drop logit starts at $\sigma(D)=0.49$, just below the execution threshold. \\
Initial positions & front & Dispersed maxima of $p^{B}_t$ (water) or of the forecast ignition (retardant) in the baseline rollout; exclusion radius $12$ cells, jitter $2$ cells. \\
Initial headings & $\psi+\pi/2$ & Across the local wind at the release cell. \\
STE threshold $\tau_{\mathrm{STE}}$ & 0.5 & Threshold in the STE binary decision. \\
Drop threshold & 0.5 & Threshold for interpreting executed drops. \\
Wind normalization constant & 10.0 m/s & Maximum wind speed used in preprocessing and drift. \\
Reference coverage $c_{\mathrm{req}}$ & 2.7 & Assumed uniform dose-response reference, gal/100\,ft$^{2}$. It is not a verified fuel-specific NWCG prescription; the level-3 flow setting is also a scenario choice. \\
Suppression effectiveness $\eta$, water & 0.9 & Nominal value for the primary evaluations: fraction of burning probability removed by water at reference coverage $c_{\mathrm{req}}$. Assumed; informed by the estimates of \citet{loane1986aerial} for fires below $1{,}700$\,kW/m, and applied to water as well (see Appendix). \\
Suppression effectiveness $\eta$, retardant & 0.9 & Fraction of the fuel factor removed at the same assumed reference coverage. It is not a calibrated reduction of ignition probability. \\
Design effectiveness $\eta^{\mathrm{d}}$, water / retardant & 0.9 / 0.9 & Values for the primary schedules. Both settings also use retardant/water design pairs $0.85/0.75$ and $0.80/0.70$, evaluated at $0.9/0.9$ (Tables~\ref{tab:design_eta} and~\ref{tab:design_eta_b}). \\
Footprint scales $\ell,s,\sigma_{\perp}$ & --- & Segment length from the level-3 flow setting; dispersion fitted per aircraft to the drop-guide tables; Table~\ref{tab:footprint_calibration}. \\
Deposit window half-width & $37$ cells & Half the longest segment plus five times the largest cell-corrected dispersion, rounded up. \\
\bottomrule
\end{tabular}
\end{table*}

\begin{table*}[htp!]
\centering
\footnotesize
\caption{Loss-function weights, as archived with the reported solutions, for Setting A (total area) and Setting B (protected regions). The $\lambda$ values govern the first optimization stage; the remaining rows govern the second. In Setting A the second stage constrains burn and terminal extent; in Setting B it constrains protected-region exposure only. The backward elimination constrains selected raw losses to $\max(\rho\ell_k^{\mathrm{ref}},10^{-4})$ while removing drops.}
\label{tab:loss_weights}
\begin{tabular}{p{3.3cm}p{0.9cm}p{0.9cm}p{6.2cm}}
\toprule
\textbf{Loss term} & \textbf{A} & \textbf{B} & \textbf{Description} \\
\midrule
Burn loss $\lambda_{\mathrm{burn}}$ & 50 & 0 & Spatio-temporal fire exposure over the whole domain. \\
Final-fire loss $\lambda_{\mathrm{final}}$ & 50 & 1 & Penalizes terminal fire extent; a secondary objective in B. \\
Protection loss $\lambda_{\mathrm{prot}}$ & 0 & 100 & Fire exposure inside the protected regions. \\
Budget loss $\lambda_{\mathrm{budget}}$ & $10^{-6}$ & 0 & Deactivates non-contributing drops in the first stage (see Section~\ref{sec:scope_objective}); B inherits A's schedule. \\
Slack factor $\rho$ & 1.02 & 1.02 & Allowed degradation in the constrained metrics during the elimination. \\
Threshold floor & $10^{-4}$ & $10^{-4}$ & Absolute floor on the constraint thresholds. \\
Initial batch size & \multicolumn{2}{c}{32} & Drops removed together in the first trial of a pass. \\
Maximum batch size & \multicolumn{2}{c}{256} & Cap on the batch size after successive accepted batches. \\
Rollout allowance & \multicolumn{2}{c}{40{,}000} & Rollouts after which the stage stops; not reached in either setting. \\
\bottomrule
\end{tabular}
\end{table*}

\begin{table*}[htp!]
\centering
\footnotesize
\caption{Protected regions used in the Bear Fire experiment, as archived with the reported solution. Each region is an axis-aligned rectangle of the $619\times748$ raster, given as half-open row and column intervals after clipping to the grid. Areas use the $30\,\mathrm{m}$ cell size ($0.09$\,ha per cell). The three regions are disjoint and together cover $65.2\%$ of the modelled domain. In $\mathcal{L}_{\mathrm{prot}}$ each region is normalized by its own area before the regions are averaged, so the three contribute equally despite their different sizes.}
\label{tab:protected_regions}
\begin{tabular}{lccrrr}
\toprule
\textbf{Region} & \textbf{Rows $[y_0,y_1)$} & \textbf{Columns $[x_0,x_1)$} & \textbf{Cells} & \textbf{Area (ha)} & \textbf{Domain (\%)} \\
\midrule
$R_1$ & $[0, 200)$   & $[0, 744)$   & 148{,}800 & 13{,}392 & 32.1 \\
$R_2$ & $[201, 549)$ & $[450, 744)$ & 102{,}312 & 9{,}208  & 22.1 \\
$R_3$ & $[550, 618)$ & $[0, 744)$   & 50{,}592  & 4{,}553  & 10.9 \\
\midrule
Union & --- & --- & 301{,}704 & 27{,}153 & 65.2 \\
\bottomrule
\end{tabular}
\end{table*}

\begin{table*}[htp!]
\centering
\footnotesize
\caption{Aircraft parameters used by the intervention model. Turnaround entries are assumed physical cycle times. The implemented gate enforces minimum spacing $(\lceil\tau_a^{\mathrm{turn}}/h\rceil+1)h$, $h=0.48$\,h: $1.44$\,h for the S-2T, AT-802F and C-130; $1.92$\,h for the BAe-146 and MD-87; $2.40$\,h for the DC-10; $2.88$\,h for the 747; and $0.96$\,h for the CL-415. These conservative discrete cycles determine simulated capacity.}
\label{tab:aircraft_params}
\setlength{\tabcolsep}{4pt}
\begin{tabular}{p{3.4cm} p{1.0cm} p{1.4cm} p{1.3cm} p{1.6cm} p{1.1cm} p{1.1cm} p{1.2cm}}
\toprule
\textbf{Aircraft type (grounded days)} & \textbf{Count} & \textbf{Material} & \textbf{Payload (gal)} & \textbf{Turn-around (h)} & \textbf{Speed (m/s)} & \textbf{Drop dur. (s)} & \textbf{Drop height (m)} \\
\midrule
Grumman S-2T Turbo Tracker (2,3) & 6 & Ret & 1200 & 0.50 & 66.4 & 3.5 & 56.1 \\
Air Tractor AT-802F Fire Boss (2,3) & 2 & Ret & 820 & 0.60 & 54.0 & 3.0 & 18.3 \\
BAe-146 / Avro RJ85 (2,3,4) & 4 & Ret & 3000 & 1.10 & 64.3 & 5.0 & 45.7 \\
McDonnell Douglas MD-87 (2,3,4) & 2 & Ret & 3000 & 1.10 & 70.7 & 5.5 & 45.7 \\
Lockheed C-130 Hercules MAFFS (2,3,4) & 2 & Ret & 3000 & 0.90 & 61.7 & 4.5 & 45.7 \\
McDonnell Douglas DC-10-30 VLAT (2,3,4) & 2 & Ret & 9400 & 1.50 & 77.2 & 8.0 & 76.2 \\
Boeing 747-400 Global Supertanker (2,3,4) & 1 & Ret & 19200 & 2.20 & 77.2 & 6.0 & 76.2 \\
Canadair CL-415 Super Scooper (2,3) & 2 & Water & 1621 & 0.18 & 56.6 & 3.0 & 38.1 \\
\bottomrule
\end{tabular}
\end{table*}

\paragraph{Calibration of the drop footprints.}
The footprint of Section~\ref{sec:step2} has three scales per aircraft: the segment length $\ell=vP/q$, fixed by the payload $P$, the flow rate $q$ and the airspeed $v$; and the two dispersion scales $s$ and $\sigma_{\perp}$. We calibrate the dispersion scales from the Forest Service drop guides~\citep{lovellette2006dropguide,solarz2000at802f,solarz2000electra}, which collect the deposit of test drops in a grid of cups and tabulate, for each coverage level, the longest line obtained over the tested flow settings together with the flow rate that produced it. \emph{Coverage level} is deposited volume in gallons per $100$\,ft$^{2}$; operational recommendations depend on fuel and fire behavior~\citep{nwcg_coverage_level,nwcg2025irpg}. The guides describe these tables as an ideal case; no single drop attains the lines of two rows, because each row has its own flow setting.

In the model, the centre-line coverage of a drop at flow rate $q$ is
\[
C(u)=\frac{9.290304\,(q/v)}{\sqrt{2\pi}\,\sigma_{\perp}}
\left[\Phi\!\left(\frac{u+\ell/2}{s}\right)-\Phi\!\left(\frac{u-\ell/2}{s}\right)\right],
\]
with $\Phi$ the standard normal distribution function. Lengths are in metres; the factor $9.290304$ converts gal/m$^2$ to gal/100\,ft$^2$. The line at level $c$ is the set $\{u: C(u)\ge c\}$. Each table row $(q_k,c_k,L_k)$ is therefore a prediction of the model, and the two dispersion scales of an aircraft and material are obtained by least squares on $\log L_k$ over the rows at levels $1$ to $6$; the rows at level $0.5$, which the guides mark as normalized, and at levels $8$ and $10$, which are lines of a few tens of feet, are excluded. Table~\ref{tab:footprint_fit} compares the published and the fitted line lengths. The residuals are within $12\%$ for the S-2T with retardant, within $15\%$ for the S-2T with water and within $14\%$ for the AT-802F. For the Electra, whose tables are for $1{,}500$\,gal drops, the level-$2$ row is excluded because the table marks a drop that used less retardant and the row is inconsistent with the level-$1$ row at the same flow rate; the remaining residuals are within $20\%$. For comparison, the physically based drop model of \citet{amorim2011validation} reproduces $78\%$ of the measured line lengths within $10\%$.

The guides use the desired coverage level to select a flow setting. For the reported scenario we choose their level-$3$ settings, consistent with the assumed reference coverage of $2.7$\,gal/100\,ft$^{2}$ (Table~\ref{tab:opt_params}), without claiming that this is the operational prescription for the domain. The calibrated aircraft use the flow rate of the corresponding level-$3$ row: $430$\,gal/s for the S-2T, $277$\,gal/s for the AT-802F and $698$\,gal/s for the Electra, at the airspeeds of the corresponding tests. The airspeed enters only through the mass per unit length $q/v$, which is $6.5$\,gal/m for the S-2T and the AT-802F and $10.5$\,gal/m for the Electra. Table~\ref{tab:footprint_calibration} lists the resulting scales. Aircraft without a drop guide use an analogue with the same mass per unit length and dispersion, so that the segment length scales with the payload. The BAe-146 and the MD-87 use the Electra, whose RADS~II constant-flow tank has the same $3{,}000$\,gal capacity. The DC-10 and the 747 use the Electra as well, at $6.3$ and $12.8$ times the $1{,}500$\,gal of the calibration drops, so conclusions that rely on these two footprints carry greater uncertainty. The CL-415 uses the S-2T water pattern. The C-130 with MAFFS uses its published pattern of $3{,}000$\,gal over a quarter mile by $100$\,ft~\citep{usfs_maffs}, taken as $\ell+4s$ by $4\sigma_{\perp}$ with the dispersion of the Electra.

Two properties of this calibration deserve mention. The flow-rate dependence of the line length is explicit, so the fit uses every row of a table rather than rows from different settings. And the load enters only through $\ell$, so the tables of the Electra, which are for $1{,}500$\,gal drops, transfer to its $3{,}000$\,gal capacity without further assumption. The remaining limitation is resolution: a line $4\sigma_{\perp}=17$ to $30$\,m wide is narrower than a $30$\,m cell, and applying the diluted dose uniformly across the cell is not equivalent to treating only part of it under a nonlinear response. We express suppression through the effectiveness at reference coverage, $\eta$, so that the assumed suppression response is stated in operational coverage units.

\begin{table}[htbp]
\centering
\caption{Published and fitted line lengths (ft) per coverage level (CL, gal/100\,ft$^{2}$), with the flow rate $q$ (gal/s) of the published row. Sources: \citet{lovellette2006dropguide} (S-2T, $1{,}200$\,gal, $129$\,kn for retardant and $118$\,kn for water), \citet{solarz2000at802f} (AT-802F, $800$\,gal, $82.5$\,kn), \citet{solarz2000electra} (Electra, $1{,}500$\,gal, $129$\,kn). Fitted dispersion: S-2T retardant $s=17.9$\,m, $\sigma_{\perp}=5.03$\,m; S-2T water $36.0$, $6.74$; AT-802F retardant $31.9$, $4.22$; Electra retardant $20.1$, $6.25$.}
\label{tab:footprint_fit}
\footnotesize
\setlength{\tabcolsep}{4pt}
\begin{tabular}{lrrrr|lrrrr}
\toprule
Aircraft, CL & $q$ & Published & Fitted & Error & Aircraft, CL & $q$ & Published & Fitted & Error \\
\midrule
S-2T ret., 1 & 180 & 1,644 & 1,451 & $-12\%$ & AT-802F ret., 1 & 77 & 1,536 & 1,379 & $-10\%$ \\
S-2T ret., 2 & 180 & 960 & 960 & $0\%$ & AT-802F ret., 2 & 140 & 608 & 692 & $+14\%$ \\
S-2T ret., 3 & 430 & 595 & 569 & $-4\%$ & AT-802F ret., 3 & 277 & 423 & 390 & $-8\%$ \\
S-2T ret., 4 & 650 & 420 & 386 & $-8\%$ & AT-802F ret., 4 & 405 & 274 & 285 & $+4\%$ \\
S-2T ret., 6 & 650 & 285 & 290 & $+2\%$ & AT-802F ret., 6 & 450 & 156 & 155 & $-1\%$ \\
\midrule
S-2T water, 1 & 300 & 930 & 876 & $-6\%$ & Electra ret., 1 & 261 & 1,412 & 1,275 & $-10\%$ \\
S-2T water, 2 & 640 & 550 & 468 & $-15\%$ & Electra ret., 3 & 698 & 408 & 474 & $+16\%$ \\
S-2T water, 3 & 640 & 360 & 363 & $+1\%$ & Electra ret., 4 & 698 & 349 & 420 & $+20\%$ \\
S-2T water, 4 & 640 & 220 & 254 & $+15\%$ & Electra ret., 6 & 698 & 238 & 233 & $-2\%$ \\
S-2T water, 6 & 955 & 65 & 64 & $-1\%$ & & & & & \\
\bottomrule
\end{tabular}
\end{table}

\begin{table}[htbp]
\centering
\caption{Footprint scales of the fleet: segment length $\ell=vP/q$, dispersion $s$ (along the track) and $\sigma_{\perp}$ (across), mass per unit length $q/v$ and the plateau coverage on the centre line, $9.290304(q/v)/(\sqrt{2\pi}\sigma_{\perp})$. Provenance: M, calibrated from the aircraft's own drop guide; A, analogue (same $q/v$ and dispersion as the aircraft in parentheses); P, published pattern extent.}
\label{tab:footprint_calibration}
\footnotesize
\setlength{\tabcolsep}{4pt}
\begin{tabular}{llrrrrrrl}
\toprule
Aircraft & Material & Payload & $\ell$ & $s$ & $\sigma_{\perp}$ & $q/v$ & Plateau & Prov. \\
 & & (gal) & (m) & (m) & (m) & (gal/m) & (gal/100\,ft$^2$) & \\
\midrule
Air Tractor AT-802F & retardant & 820 & 126 & 31.9 & 4.22 & 6.5 & 5.7 & M \\
CDF S-2T Turbo Tracker & retardant & 1{,}200 & 185 & 17.9 & 5.03 & 6.5 & 4.8 & M \\
BAe-146 / Avro RJ85 & retardant & 3{,}000 & 285 & 20.1 & 6.25 & 10.5 & 6.2 & A (Electra) \\
McDonnell Douglas MD-87 & retardant & 3{,}000 & 285 & 20.1 & 6.25 & 10.5 & 6.2 & A (Electra) \\
C-130 with MAFFS & retardant & 3{,}000 & 322 & 20.1 & 7.62 & 9.3 & 4.5 & P \\
McDonnell Douglas DC-10 & retardant & 9{,}400 & 894 & 20.1 & 6.25 & 10.5 & 6.2 & A (Electra) \\
Boeing 747-400 & retardant & 19{,}200 & 1{,}825 & 20.1 & 6.25 & 10.5 & 6.2 & A (Electra) \\
Canadair CL-415 & water & 1{,}621 & 154 & 36.0 & 6.74 & 10.5 & 5.8 & A (S-2T water) \\
\bottomrule
\end{tabular}
\end{table}

\paragraph{Suppression response and the effectiveness parameter.}
Footprint calibration determines deposited coverage; the reference coverage and suppression effectiveness are separate scenario assumptions informed by, but not calibrated from, published measurements.

Because the footprint is normalized to its discrete sum, the material reaching a cell is $V\phi(x)$ gallons, where $V$ is the payload and $\phi$ the normalized footprint. A $30$\,m cell is $96.9$ units of $100$\,ft$^{2}$, so the deposit corresponds to a coverage level $c(x)$ in the units aerial suppression is specified in. We define the response as
\begin{equation}
\mathrm{supp}(x) \;=\; 1-(1-\eta)^{\,c(x)/c_{\mathrm{req}}},
\label{eq:suppression_dose}
\end{equation}
so that $\eta$ is the fraction removed from the burning probability (water) or fuel factor (retardant) when a cell receives the assumed reference coverage. Because retardant changes both spread potential and ignition gain before a nonlinear ignition update, $\eta$ is not generally the fractional reduction of ignition probability or fireline intensity. The parameter is bounded in $[0,1)$: $\eta=0$ is no effect, $\eta=0.9$ removes $90\%$ at reference coverage, and $\eta\to1$ is the limit of complete suppression, approached but never reached. Internally the surviving fraction is $\exp(-\text{dose})$ with $\text{dose}=-\ln(1-\eta)\,c/c_{\mathrm{req}}$, because the exponential form makes repeated drops compose multiplicatively, as the rollout requires. The parameterization is a restatement, not a change of dynamics.

We assume a spatially uniform reference coverage $c_{\mathrm{req}}=2.7$\,gal/100\,ft$^{2}$ for this scenario. It is not a verified NWCG prescription or a calibrated requirement for each cell. The LANDFIRE raster uses the $40$ Scott--Burgan fuel models~\citep{landfire,scott2005standard}; an operational translation from these classes to retardant coverage requires a justified mapping and fire-behavior adjustment. For example, the NWCG guidance lists longneedle conifer at coverage level $2$ and shortneedle conifer with heavy dead litter at level $4$~\citep{lovellette2006dropguide,nwcg2025irpg}. This does not support treating a TL8 assignment of level $4$ as prescribed. We therefore interpret $2.7$ as an assumed normalization for the reported experiments. It determines both dose response and the choice of level-$3$ flow settings, and changing it would require reevaluating the schedules.

\paragraph{Interpretation and selection of $\eta$.}
Larger $\eta$ gives more suppression at fixed coverage. The concave response yields diminishing returns from concentrating material on the same cell:
\begin{center}
\begin{tabular}{lcccc}
\toprule
$\eta$ & 0.50 & 0.60 & 0.70 & 0.90 \\
\midrule
at reference coverage & $50.0\%$ & $60.0\%$ & $70.0\%$ & $90.0\%$ \\
at half reference coverage & $29.3\%$ & $36.8\%$ & $45.2\%$ & $68.4\%$ \\
at twice reference coverage & $75.0\%$ & $84.0\%$ & $91.0\%$ & $99.0\%$ \\
\bottomrule
\end{tabular}
\end{center}

We assume $\eta=0.9$ for both materials. No study measures this quantity directly; the value is informed by the estimates of \citet{loane1986aerial}, reproduced by \citet{plucinski2007aerial}, of the fraction of the original fire intensity that remains after a retardant drop, as a function of the retardant depth on the ground and by fireline-intensity class. Our assumed reference coverage of $2.7$\,gal/100\,ft$^{2}$ is a depth of $1.1$\,mm. Read from their curves at that depth, the remaining fraction is approximately $0.1$ for fires below $1{,}700$\,kW/m and approximately $0.3$ for fires of $1{,}700$ to $3{,}000$\,kW/m; above $3{,}000$\,kW/m the remaining fraction does not decrease below approximately $0.16$ at any depth, and above $5{,}000$\,kW/m they report no useful retarding effect. We identify the remaining fraction with $1-\eta$. This identification is ours: their quantity is a fireline intensity ratio estimated from field observations, whereas $\eta$ acts on the burning probability or on the fuel factor of a cell. The curves have the saturating form of \eqref{eq:suppression_dose}, and laboratory fuel beds treated with a polyphosphate retardant give a fire-intensity reduction factor of $0.80$ that is constant across the intensities of the untreated front~\citep{agueda2011intensity}, which is consistent with a multiplicative survival factor.

The relevant intensity class depends on the event. The simulated window, $19$ August to $3$ September $2020$, precedes the wind event of $8$ September. In this period the reported area of the Bear Fire grew from $41$ acres on $20$ August to $12{,}154$ acres on $4$ September~\citep{inciweb2020northcomplex}, which the siege report describes as slow burning~\citep{calfire2020}. The same report states that the fire was in patrol status from $30$ August; the daily updates do not support that statement, since they record growth from $8{,}302$ to $12{,}154$ acres and containment rising from $0\%$ to $29\%$ between $30$ August and $4$ September. Over the window the ERA5-Land daily mean wind is $0.2$ to $1.3$\,m/s, the perimeter-averaged spread in the perimeter record is $25$ to $245$\,m/day, and the largest daily advance of the head, measured as the distance from the previous perimeter, is $0.1$ to $2.3$\,km. The cells that burned are mostly long-needle conifer litter (fuel model TL8), with some shrub (SH2) and timber-understory (TU1) cells. We estimate the fireline intensity with Byram's relation $I=Hw R$, where $H=18{,}000$\,kJ/kg is the heat of combustion, $w$ the fine fuel consumed per unit area and $R$ the rate of spread. With the fine-fuel load $w=1.3$\,kg/m$^{2}$ of model TL8~\citep{scott2005standard}, assumed fully consumed, these advances give head-fire intensities of approximately $100$ to $600$\,kW/m as a $24$-hour average and approximately $300$ to $1{,}900$\,kW/m over an assumed active burning period of eight to ten hours. This estimate places the event in or near the lower class of \citet{loane1986aerial}, for which the remaining fraction at $1.1$\,mm is approximately $0.1$; this is the basis of the value $0.9$. The value would not apply to the wind-driven run of $8$ September~\citep{calfire2020}, which is outside the simulated window.

Water is less effective than long-term retardant per unit depth, and we use the same value for both materials. \citet{loane1986aerial}, as reproduced by \citet{plucinski2007aerial}, give curves of the depth required to hold a fire as a function of its intensity. Read from these curves at $6$\,MW/m, the required depth is approximately $3.1$\,mm of water against $1.15$\,mm of long-term retardant in eucalypt litter, and $1.7$\,mm against $0.7$\,mm in grass, so water requires approximately $2.5$ to $2.7$ times the depth of retardant; \citet{luke1978bushfires} state that long-term retardants are about three times as efficient as plain water on burning slash. Under \eqref{eq:suppression_dose} a depth ratio $k$ gives $1-\eta_{\mathrm{w}}=(1-\eta_{\mathrm{ret}})^{1/k}$, which is $\eta_{\mathrm{w}}\approx0.57$ to $0.60$ for $k$ between $2.5$ and $2.7$ at $\eta_{\mathrm{ret}}=0.9$. We retain $\eta_{\mathrm{w}}=0.9$ as a scenario assumption. Over most of the estimated intensity range of this event, water at the assumed reference depth of $1.1$\,mm meets the depth that the same curves require, approximately $0.65$\,mm at $1$\,MW/m; the difference between the materials is then mainly persistence, which the model represents through the one-time action of water and the persistent action of retardant. Water accounts for $56$ of $1{,}111$ drops in Setting A and $2$ of $735$ in Setting B, but counts alone do not determine its marginal contribution. The effect of lowering water effectiveness alone on the reported schedules has not been isolated.

\paragraph{Relation between $\eta$ and a drop-level success rate.}
Field studies report drop-level outcomes, not $\eta$. The study of \citet{usfs2020afue} reports, for each aircraft class, the fraction of interacting drops that achieved their stated objective: $0.85$ for helicopters and $0.81$ for scoopers, both dropping water, and $0.70$ for airtankers dropping long-term retardant. These fractions aggregate placement, coverage, timing, fire behavior, and tactical objectives. They do not identify the cell-level burning-probability or fuel-factor multiplier in this model, and we do not use them to estimate or bound $\eta$.

The value $\eta=0.9$ does not resolve the sub-cell treatment approximation described above, and the estimates that inform it are for Australian fuels. Reported reductions depend on it; Section~\ref{sec:limitations} states the consequence.

\paragraph{Preservation of the probability simplex.}
Without active clipping, the updates conserve probability: water transfers $p^B$ to $p^U$, and the fire transition transfers $p^U$ to $p^B$ and $p^B$ to $p^R$ in equal and opposite amounts. The implementation clips each component separately, so numerical checks remain useful. In the final transition-error-only ensemble, the largest normalization deviation recorded at daily boundaries across the baseline and six schedules is $6.1\times10^{-6}$ when rounded upward. No negative probability, decrease in burned probability, or burning on a nonburnable cell was recorded. The state-only and joint ensembles also pass the daily-boundary checks before and after sampling, including preservation of already burned cells. These are numerical checks at the recorded boundaries; the transition equations establish the allowed transfers within each day.

To summarize, the intervention appendix defines the decisions and rollout equations. It also lists the feasibility constraints and numerical settings used in the case study.

\section{Reference Planner Specifications}
\label{app:planners}

This appendix specifies the benchmark planners. They use the same frozen simulator and intervention rollout as the gradient optimizer.

\subsection{Common protocol}

\paragraph{One shared objective.}
Every planner is scored by, and where applicable optimizes, the same scalar. Let $\mathbf{d}$ denote the drop logits, $\boldsymbol{\ell}$ the pose logits, and $N_{\mathrm{drop}}$ the number of executed drops. Then
\begin{equation}
J(\mathbf{d},\boldsymbol{\ell})
= \underbrace{w_{\mathrm{b}}\overline{P^{\mathrm{fire}}}
+ w_{\mathrm{f}}\overline{P^{\mathrm{fire}}_T}
+ w_{\mathrm{p}}\overline{P^{\mathrm{fire}}_{\mathcal{R}}}
+ w_{\mathrm{g}} N_{\mathrm{drop}}}_{\text{the objective of Appendix~\ref{app:Intervention Optimization}}}
\;+\;
\lambda\,\frac{\max(0,\;N_{\mathrm{drop}}-B)}{B},
\label{eq:shared_objective}
\end{equation}
where $B$ is $1{,}111$ in Setting A and $735$ in Setting B, and $\lambda=10$. The first four terms use Table~\ref{tab:loss_weights}; the last penalizes excess drops through the execution STE. One implementation supplies table scores, black-box evaluations, and the gradient planner's differentiated objective.

\paragraph{Matched budgets.}
All planners share the fleet, the grounding calendar, the turnaround times, and the budget $B$. Search-based planners receive $6{,}000$ simulator rollouts. One Adam epoch is a forward rollout plus a reverse sweep; Section~\ref{sec:scaling} measures that at $2.05$ rollouts on average, and it is charged as three, so the gradient planner is allowed $2{,}000$ epochs.

\paragraph{Planner-specific initialization.}
The derivative-free planners initialize activation near the budget and poses at the baseline's per-day fire-front centroid. Activation thresholds are calibrated to an executed count within five drops of $B$. Random schedules use the same budget calibration with sampled poses; the tactical planner builds from forecasts, and greedy starts with no drops. The budgeted gradient planner uses the Stage~1 initialization of Section~\ref{sec:opt}, with every drop logit just below the execution threshold, so that it starts with no executed drops and the fire gradient activates drops under the budget penalty.

\paragraph{The execution rule does not depend on the fire state.}
Whether aircraft $a$ drops at step $t$ depends only on the logits, the grounding calendar, and the turnaround counters:
\begin{equation}
d_{t,a} = \mathds{1}\!\left[\sigma(\mathbf{d}_{t,a})\,\mathds{1}[c_{t,a}=0]\,(1-g_{t,a}) > \tau\right],
\qquad
c_{t+1,a} = d_{t,a}\,\Delta_a + (1-d_{t,a})\max(c_{t,a}-1,0),
\label{eq:gate}
\end{equation}
Here $\tau$ is the execution threshold, $c_{t,a}$ is a cooldown counter, and $g_{t,a}=1$ indicates grounding. With $h=24/S_{\mathrm{day}}$ hours, the reset value is $\Delta_a=\lceil\tau_a^{\mathrm{turn}}/h\rceil$. Because availability is checked before decrementing, a drop at step $t$ next permits execution at $t+\Delta_a+1$. This is the conservative rule used by every reported planner; it is stricter than enforcing rounded-up turnaround alone. These counter and grounding conventions differ from the binary availability gates above. Execution is independent of the fire state, so budget calibration and the feasibility audit can reconstruct it without a fire rollout. The reconstruction was checked against the simulator's schedule and counters.

\subsection{Random feasible planners}

Each (time, aircraft) pair receives a uniform random priority. Threshold bisection selects a schedule with executed count within five of $B$. The \textit{uniform} variant samples locations across the raster; the \textit{burn-footprint} variant samples cells whose baseline burning probability ever exceeds $0.05$. Both sample orientation uniformly on $[0,\pi)$. The latter controls for coarse knowledge of where the fire spreads without matching drop and fire-arrival times.

\subsection{Tactical heuristic}

This planner follows aerial-suppression doctrine and performs no optimization. Let $b_t$ be the burning field of the current rollout. At each step, aircraft that are off turnaround and not grounded are assigned targets:
\begin{itemize}
\item \textbf{Direct attack (water).} Guidance field $b_t$: the strongest active flaming front.
\item \textbf{Indirect attack (retardant).} Guidance field $\max(0,\,b_{t+h}-b_t)$ with lead $h=25$ steps (half a day): cells the simulator forecasts will burn but that are not yet burning, i.e. line laid ahead of the advancing head.
\end{itemize}
Targets are drawn as dispersed local maxima by greedy non-maximum suppression with a $12$-cell exclusion radius, so a simultaneous wave forms a line rather than concentrating on one cell, with a two-cell aim-point jitter representing dispersion. Each footprint is laid \emph{across} the local spread direction. The footprint's long axis lies along $(\cos\theta,-\sin\theta)$ in $(\mathrm{d}x,\mathrm{d}y)$ grid coordinates. A wind direction $\psi$ points along $(\cos\psi,-\sin\psi)$ in the same north-up frame. Blocking the advance therefore requires
\begin{equation}
\theta = \left(\psi+\tfrac{\pi}{2}\right) \bmod \pi .
\end{equation}
Aircraft are allocated earliest-feasible, with turnaround advanced exactly as in \eqref{eq:gate}. One pass therefore consists of a forecast, obtained by simulating the current schedule, followed by one sweep of placements over the horizon. Because the placed drops change the fire, the planner re-simulates under the schedule it has just produced and sweeps again; it runs three passes and keeps the best of the three schedules, a fourth pass changing the schedule by less than one drop. This is the open-loop analogue of a tactical re-planning cycle.

\subsection{Greedy marginal-benefit planner}

Starting from no drops, greedy proceeds in rounds. It first simulates the current schedule $u$ and draws up to $32$ feasible candidate drops from the predicted front, using material-specific guidance as above. Each candidate specifies a time, aircraft, and pose. It evaluates every candidate separately against the same $u$, sorts by objective, and commits improving candidates up to one per aircraft and at most $21$ per round. The combined schedule is then re-evaluated.

Standard greedy adds only the best candidate~\citep{nemhauser1978greedy}; at $32$ evaluations per addition, filling the $1{,}111$-drop budget would require about $36{,}000$ rollouts. Batching reduces that cost but does not guarantee that joint improvement equals the sum, or even the largest, of the isolated gains. Fire dynamics and cooldown changes couple the additions. The one-per-aircraft limit avoids conflicts between new candidates for the same aircraft; it does not prevent a new drop from changing that aircraft's later execution schedule.

Candidate scores are independent and could be evaluated in parallel, although the current implementation evaluates them sequentially. A full round requires one forecast rollout, up to $32$ candidate evaluations, and a combined-schedule evaluation. The planner stops at the drop budget, evaluation allowance, or a round with no improving candidate. Setting A terminates with $1{,}103$ drops after $2{,}333$ rollouts; Setting B exhausts its $735$ drops after $1{,}532$ rollouts. The candidate count was not tuned.

\subsection{Derivative-free planners}

The derivative-free planners use $1{,}260$ daily variables: one activation and three pose variables per aircraft and day, repeated over that day's steps. Variables are bounded to $[-3,3]$. CMA-ES uses the reference implementation with diagonal covariance and default population size~\citep{ros2008sepcma}; differential evolution and particle swarm use nevergrad defaults. Diagonal covariance avoids the quadratic storage of full-covariance CMA-ES. The native $63{,}000$-variable CMA-ES run is reported separately as a control.

\subsection{Gradient planner in the benchmark harness}

The benchmark runs the gradient planner directly to measure runtime and evaluation count. It differentiates \eqref{eq:shared_objective} over all $63{,}000$ variables using Adam at learning rate $3\times10^{-3}$, the rate of the reported first stage, gradient-norm clipping at $1.0$, and drop-logit clamping at $\pm3$. The differentiated budget penalty penalizes improvements in fire performance obtained through excess drops. The reported two-stage schedule uses a separate first-stage run followed by pruning and is listed outside the shared allowance.

\end{appendices}

\bibliographystyle{sn-mathphys-num}
\bibliography{bibliography/refs}

\end{document}